\documentclass[11pt]{article}
\usepackage{comment,soul}
\usepackage[hidelinks]{hyperref}
\usepackage{enumitem}
\usepackage{amsmath,amssymb,epsf,cite,graphicx,subfigure}
\usepackage{amsmath,braket,tikzsymbols}
\usepackage[bbgreekl]{mathbbol}
\usepackage{comment}
\usepackage[export]{adjustbox}
\usepackage{dsfont}
\usepackage{faktor}
\usepackage{mathrsfs}
\usepackage{float}

\numberwithin{equation}{section}

\usepackage{graphicx}
\usepackage{amsmath}
\DeclareMathOperator{\sech}{sech}
\usepackage{amssymb}
\usepackage{amsthm}
\usepackage{tikz}
\usepackage{mathrsfs}
\usepackage{multirow}
\usepackage{scalerel}
\usepackage{mathtools}
\usepackage{textcomp}
\usepackage{dsfont}
\usepackage{blkarray}
\usepackage{bm}
\usepackage{todonotes}
\usepackage{hyperref}
\usepackage[normalem]{ulem}

\usepackage{pgfplots}
\usepackage{tikz}
\usetikzlibrary{calc}
\usepackage{tikz-3dplot}
\usetikzlibrary{patterns}
\usetikzlibrary{knots}
\usetikzlibrary{decorations}
\usetikzlibrary{decorations.pathreplacing}
\usetikzlibrary{calc,math,angles,quotes}
\usetikzlibrary{decorations.markings,arrows.meta}

\allowdisplaybreaks[4]
\definecolor{darkyellow}{rgb}{0.5, 0.5, 0.0}
\definecolor{darkpurple}{rgb}{0.5, 0.2, 0.8}
\definecolor{darkblue}{rgb}{0.0, 0.0, 0.8}
\definecolor{darkgreen}{rgb}{0.0, 0.4, 0.0}
\definecolor{darkred}{rgb}{0.5, 0.0, 0.0}
\hypersetup{
    linktocpage,
     colorlinks,
     citecolor=darkgreen,
     linkcolor= darkgreen,
     urlcolor=darkgreen
}

\newcommand{\Rb}{\overline{R}}

\definecolor{col1}{rgb}{0.81, 0.85, 0.91}
\definecolor{col2}{rgb}{0.96, 0.88, 0.74}
\definecolor{col3}{rgb}{0.87, 0.91, 0.76}
\definecolor{col4}{rgb}{0.98, 0.82, 0.76}
\definecolor{col5}{rgb}{0.86, 0.84, 0.91}
\definecolor{col6}{rgb}{0.93, 0.83, 0.73}

\newcommand{\Ndelta}{N_X[x]}
\newcommand{\Nxzero}{N_{x_0}[x]}
\newcommand{\Ndeltanox}{N_X}
\newcommand{\Ndeltaphi}{N_{d}[\phi]}

\newcommand{\stkout}[1]{\ifmmode\text{\sout{\ensuremath{#1}}}\else\sout{#1}\fi}

\begin{document}
\baselineskip=15.5pt
\pagestyle{plain}
\setcounter{page}{1}

\begin{center}
{\LARGE \bf  The Collective Coordinate Fix}
\vskip 1cm

\textbf{Arindam Bhattacharya,$^{1}$ Jordan Cotler,$^{1,2}$ Aur\'{e}lien Dersy,$^{1}$ Matthew D. Schwartz$^{1}$}

\vspace{0.5cm}

{\it ${}^1$ Department of Physics, Harvard University, Cambridge, MA 02138, USA \\}
{\it ${}^2$ Society of Fellows, Harvard University, Cambridge, MA 02138, USA\\}

\vspace{0.3cm}

{\tt arindamb@g.harvard.edu, jcotler@fas.harvard.edu, \\ adersy@g.harvard.edu, schwartz@g.harvard.edu\\}

\medskip

\end{center}

\vskip1cm

\begin{center}
{\bf Abstract}
\end{center}
\hspace{.3cm} 
Collective coordinates are frequently employed in path integrals to manage divergences caused by fluctuations around saddle points that align with classical symmetries.  These coordinates parameterize a manifold of zero modes and more broadly provide judicious coordinates on the space of fields.  However, changing from local coordinates around a saddle point to more global collective coordinates is remarkably subtle. The main complication is that the mapping from local coordinates to collective coordinates is generically multi-valued.  Consequently one is forced to either restrict the domain of path integral in a delicate way, or otherwise correct for the multi-valuedness by dividing the path integral by certain intersection numbers.  We provide a careful treatment of how to fix collective coordinates while accounting for these intersection numbers, and then demonstrate the importance of the fix for free theories.  We also provide a detailed study of the fix for interacting theories and show that the contributions of higher intersections to the path integral can be non-perturbatively suppressed.  Using a variety of examples ranging from single-particle quantum mechanics to quantum field theory, we explain and resolve various pitfalls in the implementation of collective coordinates.

\newpage
\tableofcontents

\newpage

\section{Introduction}
One of the great advantages of the path integral for a quantum theory is that it allows for expansions around saddle points other than the free theory, providing access to non-perturbative physics. For example, the lifetime of the Standard Model can be computed by expanding the path integral around a non-trivial configuration of the Higgs field~\cite{Frampton:1976kf,Coleman:1977py,Callan:1977pt,Isidori:2001bm
,Andreassen:2016cff, Andreassen:2016cvx, Andreassen:2017rzq}.  In many field theories (including the Standard Model) there are families of equivalent non-trivial saddles related by symmetries, such as translations, dilatations, or internal symmetries.  Expanding around any one of these saddles leads to a divergent 1-loop saddle point calculation because the zero eigenvalues from the flat directions make the functional determinant vanish. Such infinities are usually handled by introducing collective coordinates along the manifold of zero modes, effectively trading local coordinates around a single saddle for more global coordinates around the entire manifold of zero modes \cite{Gervais:1974dc, Bernard:1979qt, tHooft:1976snw}. Then the integral over the zero mode manifold can be handled non-perturbatively, giving a factor of the volume of the symmetry group.  The remaining path integral can then be evaluated using a saddle-point approximation, and can be regularized to be finite.  While this procedure gives sensible answers in some situations where it can be checked (for example, the energy splitting in a symmetric double-well potential can be alternatively computed with the WKB approximation \cite{AIHPA_1983__39_3_211_0,DELABAERE1997180,10.1063/1.1767988,Dunne:2014bca,PhysRevA.86.012106,10.1119/1.19458}), the procedure also happens to give very wrong answers in other situations, such as in free theories. 

To see the problem clearly, consider in more detail the decay rate calculation in a quantum field theory with action $S = \int d^4 x \left(\frac{1}{2}\,\partial_{\mu} \phi\,\partial^\mu \phi + \lambda \phi^4 \right)$ with $\lambda < 0$. As we will review in Section~\ref{sec:qft}, the decay rate involves an expansion around a non-trivial instanton background $\phi_b(x)$ normalized by an expansion around the trivial vacuum $\phi(x) = 0$. The decay rate $\Gamma$ can be written as \cite{Andreassen:2016cvx,Andreassen:2016cff,Andreassen:2017rzq
}
\begin{equation}
\Gamma = \lim_{T \rightarrow \infty} \frac{1}{T} \text{Im} \frac{\int_{C_b}\mathcal{D} \phi \,e^{- S[\phi]}}{\int_{C_{\text{FV}}} \mathcal{D} \phi \,e^{-S_0[\phi]}}\,, \label{FVdecay1}
\end{equation}
where $S_0[\phi] = \int d^4x\left( \frac{1}{2} \partial_\mu \phi \partial^\mu \phi\right)$ is the free action. Here $C_b$ is a path of steepest descent in the interacting theory passing through $\phi_b(x)$ and $C_{\text{FV}}$ is the path of steepest descent in the free theory passing though the free vacuum $\phi(x) = 0$. To evaluate the numerator in the saddle point approximation, we expand
\begin{equation}
  \phi (x) = \phi_b (x) + \sum_{m=0}^\infty a_m \phi_m (x)\,,
\end{equation}
where 
$\{\phi_m (x)\}$ forms a basis of normal modes around the bounce $\phi_b(x)$. Five of these modes, corresponding to four translations and one dilatation mode, have zero eigenvalue with respect to the operator $L''[\phi_b]$, where $L$ is the Lagrangian, and so a na\"{i}ve saddle point approach gives infinite contributions to the numerator of Eq.~\eqref{FVdecay1}.  A more careful treatment of steepest descent is required to resolve these infinities.  To do so, we can change to a collective
coordinate basis
\begin{equation}
  \phi (x) = \phi_b (x + \xi, R) + \sum_{m \geq 5} \alpha_m \phi_m (x + \xi, R)\,,
  \label{basisshift}
\end{equation}
where the five zero-mode coordinates $\alpha_m$ for $m = 0,1,2,3,4$ are traded for five collective coordinates $\xi^{\mu}$ and $R$. 
The integral over $\xi^\mu$ then gives a factor of the spacetime volume $T\mathcal{V}$ for the numerator of Eq.~\eqref{FVdecay1}, as expected for a decay rate in spacetime. However, a complication arises if we use the same exact basis of Eq.~\eqref{basisshift}, which happens to also diagonalize normal modes in the free theory,
for the denominator of Eq.~\eqref{FVdecay1}. In particular, the same $T\mathcal{V}$ factor apparently arises, which is unexpected.  In this paper, we identify the cause of this unexpected factor, and show that the appropriate treatment of collective coordinates yields sensible results for both the numerator and denominator of Eq.~\eqref{FVdecay1}.  We more broadly study general coordinates on field space in which similar subtleties arise.

To understand the source of the subtleties, let us recall how $\delta$-functions can be used to change variables in integrals.  In a finite-dimensional integral $\int_{\Omega} d^d x \, A(x)$ over $x \in \Omega \subseteq \mathbb{R}^d$, we can implement the change of variables $x = f(y)$ using the resolution of the identity
\begin{align}
\label{E:res1}
1 = \int_{f^{-1}(\Omega)}\! d^d y \, \left|\frac{\partial f(y)}{\partial y}\right|\,\delta^d(f(y) - x)\,,
\end{align}
so that
\begin{align}
\int_\Omega d^d x\,A(x) &= \int_\Omega d^d x \int_{f^{-1}(\Omega)}\! d^d y \, \left|\frac{\partial f(y)}{\partial y}\right|\,\delta^d(f(y) - x)\,A(x) \\
&=  \int_{f^{-1}(\Omega)}\! d^d y \, \left|\frac{\partial f(y)}{\partial y}\right|\,A(f(y))\,.
\end{align}
The above can be regarded as a trick to find the right Jacobian factor and domain of integration.  Crucially, we notice that the above assumes that $f$ is an invertible function on the domain of integration.  But what if $f$ multiply-covers the domain of integration?  A simple example is the 1-dimensional integral $\int_0^\infty dx \, A(x)$ with the change of variables $x = y^2$.  In this setting the salient resolution of the identity is
\begin{align}
1 = \frac{1}{2} \int_{-\infty}^\infty dy\,|2y|\,\delta(y^2 - x)\,.
\end{align}
Here, we need the factor of $\frac{1}{2}$ out front because the map $f(y) = y^2$ taking $\mathbb{R} \to \mathbb{R}_{\geq 0}$ doubly covers the positive real line.  Going back to Eq.~\eqref{E:res1}, if $f(y) = x$ has $N_f[x]$ solutions for fixed $x$, then we need to correct Eq.~\eqref{E:res1} by
\begin{align}
\label{E:res2}
1 = \frac{1}{N_f[x]}\int_{f^{-1}(\Omega)}\! d^d y \, \left|\frac{\partial f(y)}{\partial y}\right|\,\delta^d(f(y) - x)\,,
\end{align}
where we call $N_f[x]$ the \textit{intersection number}.  A similar subtlety arises in gauge theories where there can be multiple solutions to a local gauge-fixing condition, and so we must introduce a multiplicative factor to correct for the so-called \textit{Gribov copies}~\cite{Gribov:1977wm}.

To see how the intersection number appears in a path integral, consider non-relativistic quantum mechanics where the path integral is over paths $x(t)$. For a time-independent Lagrangian, there will always be a flat direction: $S[x(t)] = S[x(t+t_0)]$ for any path $x(t)$ and any $t_0$. 
To change from orthonormal coordinates on path space to collective coordinates we can employ the resolution of the identity\footnote{We have used $\Big| \langle \dot{X} (t +
  t_0) | x (t) \rangle \Big| = \Big| \langle X (t +
  t_0) | \dot{x} (t) \rangle \Big|$ obtained via integration by parts in $t$.}
\begin{equation}
\label{general1}
  1 = \frac{1}{\Ndelta} \int d t_0\, \delta\Big[\langle X (t + t_0) |x (t) \rangle\Big] \,
  \Big| \langle X (t +
  t_0) | \dot{x} (t) \rangle \Big|\,.
\end{equation}
Here $\langle x|y \rangle = \int d t \,x (t) y (t)$ denotes the $L^2$ inner product, $x(t)$ is a trajectory in the domain of the path integral, $\dot{x}(t) = \partial_t x(t)$ is the time derivative, $X(t + t_0)$ is the path whose coordinate we want to exchange
for $t_0$, and $\Ndelta$ is the intersection number. 
That is, solving $\langle X (t + t_0) |x (t) \rangle = 0$ will replace the time-translation zero mode with the collective coordinate $t_0$ for any path $x(t)$.  Then the intersection number $N_X[x]$ counts the number of values of $t_0$ for which $\langle X (t + t_0) |x (t) \rangle = 0$ for a given path $x(t)$. The intersection number is a complicated integer-valued functional of the path which must be carefully taken into account. 

The textbook narrative is that a collective coordinate can be simply pulled out of the path integral giving a volume factor~\cite{Coleman:1985rnk,Kleinert:2004ev,MullerKirsten:2012wla,Marino:2015yie,Marino:2021lne}.
Unfortunately, this standard treatment does not properly account for the multi-valuedness of the coordinate change associated with going from local coordinates to collective coordinates. In this paper, we explore through examples how and when collective coordinates can be used. Our key observations are that (1) the transition to collective coordinates does not involve only an innocuous coordinate change; (2) in interacting theories, non-Gaussian terms in the action are important to suppress high-multiplicity intersections; and (3) the Jacobian factor incurred by changing to collective coordinates plays an important role in assuring the consistency of the implementation.

Our paper begins in Section~\ref{sec:general} with an overview of how to change from local coordinates to collective coordinates. We then proceed through a set of examples of increasing complexity. In Section~\ref{sec:free}, we discuss a free particle with time compactified on a circle.  For this system time-translation invariance is exact. We explicitly compute the intersection number associated with the collective coordinates and illustrate the proper treatment of the Jacobian. In Section~\ref{sec:doublewell}, we consider the symmetric double-well potential in quantum mechanics, focusing on the computation of the energy splitting between the two lowest levels. This theory has a non-trivial instanton and a basis of excitations around the instanton whose analytic form is explicitly known.  We further show that the intersection number apropos to the collective coordinates counts the number of times a given path $x(t)$ crosses $x=0$. We then 
demonstrate that paths with $\Ndelta >1$ give exponentially small contributions to the path integral in the interacting theory, but give $\mathcal{O}(1)$ contributions in the free theory. In Section~\ref{sec:qft} we consider $\lambda \phi^4$ quantum field theory and discuss collective coordinates for dilatation and translations. 
In Section~\ref{sec:conc} we conclude with a discussion.
\section{Collective coordinates: general results \label{sec:general}}
\noindent
We begin by discussing how collective coordinates are used in quantum
mechanics. \ In evaluating the path integral, we often expand around some
saddle point $x_I (t)$. This saddle can be the free solution $x = 0$, an
instanton, or simply some classical path satisfying the boundary conditions of
the path integral for the problem of interest. For concreteness, suppose we
work in Euclidean signature with a time-independent Lagrangian of the form $L
[x] = \frac{1}{2} \dot{x}^2 + V (x)$ with action $S[x] = \int dt\, L[x]$. Then we can expand a generic path in an
orthonormal basis of eigenstates of $\frac{\delta^2 S}{\delta x^2}\big|_{x = x_I}$. That is, we write
\begin{equation}
  x (t) = x_I (t) + a_0 x_0 (t) + \sum_{n=1}^{\infty}a_n x_n (t) \label{xofa}
\end{equation}
where
\begin{equation}
  [-\partial_t^2 + V'' (x_I)] \,x_n = \lambda_n x_n\,.
\end{equation}
In~\eqref{xofa} we have separated out the zero mode $x_0 (t) \propto \dot{x}_I (t)$. We can see that $\lambda_0=0$ using $\frac{\delta S}{\delta x}\big|_{x = x_I} = 0$ so that
\begin{equation}
  0 = \partial_t \left.\frac{\delta S}{\delta x(t)}\right|_{x = x_I} = \int dt'\,\left.\frac{\delta^2 S}{\delta x(t) \delta x(t')}\right|_{x = x_I} \dot{x}_I(t')\,.
\end{equation}
Then the path integral becomes, to 1-loop order in the saddle-point
approximation,
\begin{equation}
  Z 
  = \mathcal{N}\int \mathcal{D} x\ e^{- S} 
  = \mathcal{N} e^{- S [x_I]} 
  \int \frac{d a_0}{\sqrt{2\pi}}\ \prod_{n=1}^{\infty}\int \frac{d a_n}{\sqrt{2\pi}}\ e^{-
  \lambda_n a_n^2} = \mathcal{N} e^{- S [x_I]} \sqrt{\frac{1}{\lambda_0}}
  \prod_{n=1}^{\infty}\sqrt{\frac{1}{\lambda_n}} \,.
  \label{Zform}
\end{equation}
Because of the zero eigenvalue, Eq.~\eqref{Zform} is infinite. The existence
of a zero mode is because the action $S [x (t + t_0)]$ is independent of $t_0$.
We therefore would like to trade $a_0$ for $t_0$ and integrate over $t_0$
giving a factor of the volume of time $T$.

To trade $a_0$ for $t_0$ we transform from coordinates $\{ a_0, a_n \}$ to collective coordinates 
$\{ t_0, \alpha_n \}$ of the form
\begin{equation}
  x (t) = x_I (t + t_0) + \sum_{n=1}^{\infty}\alpha_n x_n (t + t_0)\,. \label{alphaform}
\end{equation}
Since $\{ \alpha_0, \alpha_n \}$ are global coordinates, for which each path in the integration domain is represented exactly once,
the zero mode can be
projected out with
\begin{equation}
  \big\langle x (t) - x_I (t + t_0) \big|x_0 (t + t_0) \big\rangle = 0
  \label{zerocondition}
\end{equation}
where $\langle x|y \rangle$ is the inner product under which the $x_n$ are orthonormal. Eq.~\eqref{zerocondition} can be solved to find a coordinate $t_0$ for a given path $x(t)$. Importantly, however, there can be multiple solutions and we denote by $N_{x_0}[x]$ the number of solutions
for $t_0$ given $x$.
The Jacobian for the coordinate change is determined by the partial derivatives
\begin{equation}
  \frac{\partial a_m}{\partial t_0} =  \big\langle
  x_m (t) \big|\partial_{t_0}  x_I (t + t_0) + \sum_{n=1}^\infty \alpha_n \partial_{t_0}x_n (t + t_0) \big\rangle, \quad
  \frac{\partial a_m}{\partial \alpha_n} = 
  \big\langle x_m (t) \big|x_n (t + t_0) \big\rangle\,.
\end{equation}
After some algebra (see Appendix B of~\cite{Andreassen:2016cvx}), this Jacobian can be simplified to
\begin{align}
  \mathcal{J} &= \big\langle x_0 (t+t_0) \big|\partial_{t_0} {x}_I (t+t_0) + \sum_{n=1}^\infty \alpha_n \partial_{t_0} {x}_n (t+t_0)\big\rangle\\*
  &=\big\langle x_0 (t) \big|\dot{x}_I (t) + \sum_{n=1}^\infty \alpha_n \dot {x}_n (t)
  \big\rangle \label{jformtext} \,.
\end{align}
Alternatively, the functional determinant can be computed as $\sqrt{\det G}$ where $G$ is the Riemannian metric on the space of fields in collective coordinates (see Appendix~\ref{app:metric}).
That the Jacobian is independent of $t_0$ is non-trivial.  Thus we have 
\begin{equation}
\boxed{
    Z = \mathcal{N} e^{- S [x_I]} \int \frac{d t_0}{\sqrt{2\pi}}\ \int\prod_{m=1}^{\infty}  \frac{d\alpha_m}{\sqrt{2\pi}}\ e^{- \lambda_m \alpha_m^2}
    \frac{1}{N_{x_0}[x]} \Big| \big\langle x_0 (t) \big| \dot{x}_I (t) + \sum_{n=1}^\infty \alpha_n
    \dot{x}_n (t) \big\rangle\Big| \\
}
  \label{Zform1}
\end{equation}
This is the formulation we should use if we want to implement the $\{ t_0, \alpha_n
\}$ coordinates as in Eq.~\eqref{alphaform}.

From Eq.~\eqref{Zform1} we can integrate in $\alpha_0$ by including $\int d
\alpha_0 \,\delta (\alpha_0)$. If we then relabel $\alpha_0 \rightarrow a_0$ and
$\alpha_n \rightarrow a_n$, the resulting expression would be exactly the same
as if we had expanded
\begin{equation}
  Z = \mathcal{N} \int d t_0 \left\{\int\mathcal{D} x\ e^{- S [x]} \frac{1}{N
  _{x_0}[x]}
  \delta\Big[\big\langle x_0 (t) \big|x (t)\big\rangle \Big] \,
  \Big| \big\langle x_0 (t) \big| \dot{x}  (t)
\big\rangle \Big| \right\} \label{Zform2}
\end{equation}
using Eq.~\eqref{xofa}. In the above equation we can no longer use $t_0$ as a
coordinate: it is a constant from the point of view of the path
integral (the part of the expression in the curly brackets). That is, we cannot evaluate the path integral expression
using the coordinates in Eq.~\eqref{alphaform}. If we want to use $t_0$ as a
coordinate, we should instead use Eq.~\eqref{Zform1}.

There was nothing essential about $x_I (t)$ or $x_0 (t)$ in our derivation.
Indeed, $x_I (t)$ does not even appear in Eq.~\eqref{Zform2}. We can replace
$x_0 (t)$ by any path $X (t)$ leading to
\begin{equation}
\boxed{
    Z = \mathcal{N} \int d t_0 \left\{ \int\mathcal{D} x\ e^{- S [x]} \frac{1}{\Ndelta}
  \,\delta\Big[\big\langle X (t) \big|x (t)\big\rangle \Big] \,
  \Big| \big\langle X (t) \big| \dot{x}  (t)
  \big\rangle \Big| \right\} \\
}
\label{Zcollmain}
\end{equation}
where $\Ndelta$ counts the number of times $\langle X(t+t_0)| x(t) \rangle$ vanishes over the integration range of $t_0$ for fixed $x(t)$. Note that the above equation can also be derived inserting the identity of Eq.~\eqref{general1} inside the path integral and using $S[x(t+t_0)]=S[x(t)]$.

To summarize, the key new element in our analysis is the intersection number $\Ndelta$. If we write a generic path $x(t)$ in an orthonormal basis, then $\Ndelta$ is the number of solutions to
\begin{equation}
  \langle X (t + t_0) |x (t) \rangle = 0 \label{Xcondition}
\end{equation}
over the range of $t_0$. Eq.~\eqref{Xcondition} allows us to solve for $t_0$ in terms of the coordinates labeling an orthonormal basis, and there may be multiple solutions.
When the collective coordinate $t_0$ is one of the coordinates, Eq.~\eqref{Xcondition} cannot be used. Instead, in collective coordinates, $\Ndelta$ counts the number of ways of constructing the same path $x(t)$ using different values of $t_0$. That is, the same function would be overcounted in the path integral in the collective coordinates if it can be described using multiple values of $t_0$, and so we must divide by a $\Ndelta$ factor to compensate. We will see examples of the use of $\Ndelta$ in the following.
 
\section{Free theory \label{sec:free}}
\noindent
As a first example, we consider a free non-relativistic quantum theory for a
particle with mass $m = 1$ and Lagrangian $L = \frac{1}{2} \dot{x}^2$. The
propagator going from $a$ to $b$ in time $T$ is
\begin{equation}
  D (a, b) = \langle b | e^{- H T} | a \rangle = \int_{- \infty}^{\infty}
  \frac{d p}{2 \pi} \,e^{i p (b - a)} e^{- \frac{p^2}{2} T} = \sqrt{\frac{1}{2
  \pi T}} \,e^{- \frac{(b - a)^2}{2 T}}
\end{equation}
and the partition function is
\begin{equation}
  Z_0 = \text{Tr} (e^{- H T}) = \int_{-\mathcal{V}/2}^{\mathcal{V}/2} d x_0 \,D (x_0, x_0) = \sqrt{\frac{1}{2
  \pi T}} \,\mathcal{V}\,, \label{Zfree}
\end{equation}
where $\mathcal{V}$ is the volume of space used the regulate this integral. To compute
the partition function with the path integral, we parameterize paths with
periodic boundary conditions in time as
\begin{equation}
  x (t) = x_0 + \sum_{n\,\text{even}} \left[a_n s_n (t) + b_n c_n (t)\right]\,,
  \label{xform1}
\end{equation}
where
\begin{equation}
  s_n (t) = \sqrt{\frac{2}{T}} \sin\!\left( \frac{n \pi \left( t + \frac{T}{2}
  \right)}{T} \right), \quad c_n (t) = \sqrt{\frac{2}{T}} \cos\!\left( \frac{n
  \pi \left( t + \frac{T}{2} \right)}{T} \right).\label{sndef}
\end{equation}
Then the partition function is given by
\begin{equation}
  Z_0 = \mathcal{N} \int \mathcal{D} x\ e^{- S [x]} = \mathcal{N} \int_{-\mathcal{V}/2}^{\mathcal{V}/2} d
  x_0 \prod_{n\ \text{even}}\left[\int_{- \infty}^{\infty} \frac{d a_n}{\sqrt{2\pi}} e^{- \frac{1}{2} \lambda_n a_n^2} \int_{-
  \infty}^{\infty} \frac{d b_n}{\sqrt{2\pi}} e^{- \frac{1}{2} \lambda_n b_n^2}\right]
\end{equation}
\begin{equation}
  = \mathcal{N} \mathcal{V} \left[ \prod_{n\, \text{even}} \sqrt{\frac{1}{\lambda_n}}
  \right]^2
\end{equation}
where
\begin{equation}
  \lambda_n = \left( \frac{n \pi}{T} \right)^2
\end{equation}
are the eigenvalues of the fluctuations under $\left.\frac{\delta^2 S}{\delta x(t)\, \delta x(t')}\right|_{x = 0} = - \delta(t-t')\,\partial_t^2$. We
then match on to Eq.~\eqref{Zfree} with
\begin{equation}
  \mathcal{N} = \sqrt{\frac{1}{2 \pi T}} \left[ \prod_{n\,\text{even}}
  \frac{T}{n \pi} \right]^{- 2} \label{Npath}
\end{equation}
which fixes the normalization of the path integral. Note that $\mathcal{V}$ is only
needed to regulate the $x_0$ integral: we can take the integrals over $a_n$
and $b_n$ to go to from $-\infty$ to $\infty$ because the path integral is exponentially
suppressed at large $a_n, b_n$.

Now we want to reproduce the partition function $Z_0$ with collective coordinates, using Eq.~\eqref{Zcollmain}. We first need to choose which path $X (t)$ to project onto,
i.e.~which coordinate to swap for the collective coordinate $t_0$. The coordinate to swap should be one from a parameterization of paths where time is shifted
by $t_0$, i.e.~from
\begin{equation}
  x (t) = x_0 + \sum_{n\,\text{even}}\left(\alpha_n s_n (t + t_0) + \beta_n c_n (t + t_0)\right). \label{xform2}
\end{equation}
In the simplest case, we would swap a single coordinate, say $\beta_m$, for
$t_0$. That would mean choosing $X(t) = c_m (t)$. To change to coordinates
involving $t_0$ we have to start from coordinates not involving $t_0$, such as the
original coordinates in Eq.~\eqref{xform1}. The condition $\beta_m = 0$ is $\langle c_m (t + t_0) |x (t) \rangle = 0$ as in Eq.~\eqref{Xcondition}, which
in the original coordinates amounts to
\begin{equation}
  \tan\!\left( \frac{m \pi t_0}{T} \right) = - \frac{b_m}{a_m} \label{taneq} \,.
\end{equation}
This equation allows us to solve for $t_0$ and to transform from the $\{ a_n, b_n \}$
coordinates to the $\{ t_0, \alpha_n, \beta_n \}$ coordinates with no $\beta_m$ component.
Note that over the range $0 < t_0 \leqslant T$, Eq.~\eqref{taneq} has $m$
possible solutions so that\footnote{In this case, we can alternatively restrict
the range $0 < t_0 \leqslant \frac{T}{m}$ and take $\Ndelta = 1$, but restricting the range is untenable in more complicated examples.} $\Ndelta = m$.

The Jacobian arising from Eq.~\eqref{Zcollmain} is
\begin{equation}
  \mathcal{J} = \big\langle x (t) \big| \dot{c}_m (t) \big\rangle = \sqrt{\lambda_m} \,a_m\,,
\end{equation}
and thus in collective coordinates the partition function is
\begin{multline}
  Z_{\text{coll}} = \mathcal{N} \frac{1}{m}\int_0^{T} d t_0 \int_0^\mathcal{V} d x_0
  \prod_{ \substack{n \, \text{even} \\ n\neq m}}\left(\int_{- \infty}^{\infty} \frac{d a_n}{\sqrt{2\pi}}\ e^{- \frac{1}{2} \lambda_n a_n^2} \int_{-
  \infty}^{\infty} \frac{d b_n}{\sqrt{2\pi}} e^{- \frac{1}{2} \lambda_n b_n^2}\right) \\
  \times \int \frac{d a_m}{\sqrt{2\pi}}
  \sqrt{\lambda_m} | a_m | e^{- \frac{1}{2} a_m^2 \lambda_m} \,.
\end{multline}
Using Eq.~\eqref{Npath} the above agrees precisely with $Z_0$ in Eq.~\eqref{Zfree}.

The next simplest case is to choose $X(t)$ to be a sum of two modes instead of one mode.  Suppose we choose $X (t) = A \,c_m (t) + B \,c_k (t)$ which is a sum of cosine modes. Then the
coordinate $t_0$ would be determined from the coordinates in Eq.~\eqref{xform1} using
the condition 
\begin{equation}
  \langle X (t + t_0) |  x (t) \rangle = A \,b_m c_m (t_0) + A \,a_m
  s_m (t_0) + B \,b_k c_m (t_0) + B \,a_k s_m (t_0) = 0 \,.
\end{equation}
The number of solutions $\Ndelta$ for this equation is a
complicated function of $a_m, b_m, a_k$, and $b_k$. It turns out to be somewhat
easier to use polar coordinates. Let us write an arbitrary linear combination
of the $m$ and $k$ sines and cosine modes as
\begin{equation}
  X (t) = R_m s_m (t + \phi_m) + R_k s_k (t + \phi_k) \,.
\end{equation}
Changing to the polar coordinates $a_n = \rho_n s_n(\theta_n)$ and $b_n = \rho_n c_n(\theta_n)$ for convenience, we find the overlap
\begin{equation}
  \langle X (t + t_0) |  x (t) \rangle = R_m \rho_m s_m (t_0 +
  \theta_m + \phi_m) + R_k \rho_k s_k (t_0 + \theta_k + \phi_k)
\end{equation}
which only involves two frequencies, as expected. 

We now want to replace the $a_m, b_m, a_k, b_k$ integrals from the original partition function, which gave rise to a $\frac{1}{\lambda_m} \frac{1}{\lambda_k}$ term, with new integrals in our polar coordinates. As a cross check, we
see that before introducing collective coordinates we have
\begin{equation}
  Z_{\text{polar}} = \left( \frac{\lambda_m}{2\pi}\frac{\lambda_k}{2\pi}
  Z_0 \right) \frac{m \pi}{T} \frac{k \pi}{T} \int_0^{\infty} d \rho_m \rho_m
  e^{- \frac{1}{2} \lambda_m \rho^2_m} \int_{0}^{\infty} d \rho_k \rho_k e^{- \frac{1}{2}
  \lambda_m \rho^2_m} \int_0^{\frac{2 T}{m}} d \theta_m \int_0^{\frac{2 T}{k}}
  d \theta_k = Z_0\,,
\end{equation}
as expected. Here $\frac{\lambda_m}{2 \pi} \frac{\lambda_k}{2 \pi} Z_0$
represents the partition function where we have performed all integrals except for those over $a_m,
b_m, a_k, b_k$. The $\frac{m \pi}{T} \frac{k \pi}{T}$ factors
arise from the Jacobian induced by our polar coordinates.

\begin{figure}
    \centering
    \includegraphics[scale=0.9]{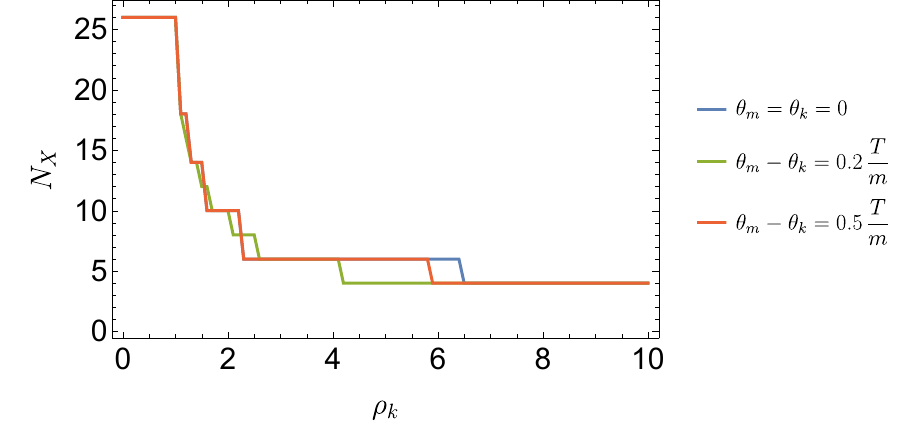}
    \caption{For the free quantum mechanical theory where the coefficient of a linear combination of two normal modes is traded for a collective coordinate, the intersection number $N_X$ cannot be neglected. $N_X$ is integer-valued. After integrating over the modulus $\rho_m$, we find that $N_X$ depends on  
    the modulus $\rho_k$ and the relative angle $\theta_m-\theta_k$. 
   Shown are the value of $N_X$ for $m=26$ and $k=4$ for a few choices of $\theta_m-\theta_k$.}
    \label{fig:thetas}
\end{figure}

Now we move to collective coordinates by inserting into the partially-integrated partition function our Eq.~\eqref{Zcollmain} which includes an integral over $t_0$, a $\delta$-function, and a Jacobian.  Doing so and then
shifting $\theta_m \rightarrow \theta_m - \phi_m$, $\theta_k \rightarrow
\theta_k - \phi_k$, rescaling $\rho_m \rightarrow \frac{\rho_m}{R_m}, \rho_k
\rightarrow \frac{\rho_k}{R_k}$, taking $\rho_k \rightarrow \rho_k \rho_m$, and
performing the $\rho_m$ integral leads to 
\begin{multline}
  Z_{\text{coll}} = Z_0\times\left( \frac{\lambda_m}{2 \pi} \frac{\lambda_k}{2 \pi} 
  \right) \frac{m \pi}{T} \frac{k \pi}{T} \int_0^{\infty} d \rho_k \frac{2
  \rho_k R_k^2 R_m^2}{(R_k^2 \lambda_m + R_m^2 \lambda_k \rho_k^2)^2}
  \int_0^{\frac{2 T}{m}} d \theta_m \int_0^{\frac{2 T}{k}} d \theta_k
\\
  \times \frac{1}{\Ndeltanox (\rho_k,\theta_k,\theta_m)} \int_0^T d t_0 \,\delta [s_m
  (\theta_m) + \rho_k s_k (\theta_k)] \left| \frac{m \pi}{T} c_m
  (\theta_m) + \rho_k  \frac{k \pi}{T} c_k (\theta_k) \right|.
  \label{Zcollform2}
\end{multline}
One can check that without the second line, we still have
$Z_{\text{coll}} = Z_0$\,. The intersection number $\Ndeltanox$
 counts the number of times the $\delta$-function fires for a given path.
We can compute
the intersection number by counting the roots of $\sin ( \pi m (t_0+\theta_m)/T) + \rho_k
\sin ( \pi k (t_0+\theta_k)/T)$ on the interval $0 < t_0 < T$.
Due to periodicity, the intersection number depends only on $\theta_k-\theta_m$.
The function for $m = 26$ and $k = 4$ with
three choices of $\theta_k-\theta_m$ is shown in Fig.~\ref{fig:thetas}.


To proceed, one can first observe that the integral is periodic in $\theta_m$ and $\theta_k$, so we can extend their limits of integration to $0<\theta_m,\theta_k<T$ and pull out an overall factor of $\frac{4}{m k}$. Then we
change variables from $(\theta_m,\theta_k)$ to 
$\theta_+ = \frac{1}{2}( \theta_m + \theta_k)$ and $\theta_- = \theta_m-\theta_k$ and integrate over $\theta_+$. In this integral, the $\delta$-function fires the
number of times given by the number of solutions of $s_m(\theta_+ + \frac{1}{2}\theta_-) + \rho_k s_k(\theta_+ -\frac{1}{2} \theta_-)$
 for $\theta_+$. But this number is identical to $\Ndeltanox(\rho_k,\theta_-)$, so that again $Z_\text{coll} = Z_0$ as expected.
One can of course integrate over  $\theta_m$ 
 or $\theta_k$ first (rather than $\theta_+$) to get the same result, but the integrals are much more difficult.
 
In the general case where $X (t)$ has components of arbitrary frequency, the
intersection number $\Ndeltanox (\rho_n, \theta_n)$ would be a function of the
entire configuration space. Thus the intersection number is generally a horrible mess, but accounting for the intersection number in the path integral is nonetheless required for collective coordinates to work. Luckily,
for some choices of $X (t)$ the intersection number has a more physical
interpretation and can be managed either analytically or approximately. A representative example is the double-well potential in quantum mechanics to which we now turn.

\section{Double well potential \label{sec:doublewell}}

For the next example, we consider the symmetric double well potential in
quantum mechanics whose Euclidean action is
\begin{equation}
  S [x] = \int_{- T / 2}^{T / 2} d t \left[  \frac{m}{2} \dot{x}^2 + V (x)
  \right], \quad V (x) = \frac{m\omega^2}{8 a^2} (x^2 - a^2)^2\,.
\end{equation}
There are two dimensionless scales in the problem: ${m a^2
\omega}/{\hbar}$ which indicates the size of the barrier and $\omega T/\hbar$ which indicates 
time. We will set $m = \hbar = 1$ but keep $a, \omega$ and $T$ to assist with
the physical interpretation of various limits.

The main quantity of interest in this theory to compute with path integral methods is the splitting of the two lowest energy eigenstates. This splitting,
and the entire spectrum, have been studied in great detail both analytically
and numerically~\cite{Brezin:1977gk,Gildener:1977sm, Bogomolny:1980ur,Zinn-Justin:1981qzi, Zinn-Justin:1982hva, Jentschura:2001kc, Zinn-Justin:2004vcw, 10.1063/1.1767988}. It is nevertheless difficult to find a rigorous analysis of
the path integral computation of the energy splitting. Typically treatments
are not only cavalier with regard to the introduction of collective coordinates, but also with regard to the associated Jacobian and finite-time regulation. We will attempt to clarify all of these issues in this section.

Since the potential $V(x)$ is symmetric, the parity operator $P$ which takes $x
\rightarrow - x$ commutes with the Hamiltonian. Thus energy eigenstates are
parity even or parity odd. In the limit $a^2 \omega \rightarrow \infty$,
the barrier becomes enormous and the spectrum is that of two harmonic
oscillators with frequency $\omega$. That is, $E_0^+ = E_0^- =
\frac{\omega}{2}$ and the higher modes are spaced by $\omega$. Expanding the
potential around a minimum and working perturbatively in the limit of large but finite $a^2 \omega$, the energies receive corrections but
the even and odd modes remain degenerate to all orders. The WKB approximation can be used to compute the energy splitting with the leading order result~\cite{10.1063/1.1767988, Zinn-Justin:2004vcw,Muller-Kirsten:2012wla,Dunne:2014bca} 
\begin{equation}
    \Delta_n^{\text{WKB}} = E_n^- -E_n^+ = \omega \frac{2}{\sqrt{2\pi}} \left(8\omega a^2\right)^{n+1/2} e^{-\frac{2\omega a^2}{3}}\,.
\end{equation}
In particular, the ground state splitting can be written as
\begin{equation}
 \Delta_0^{\mathrm{WKB}} = 4 \omega \sqrt{\frac{ \omega a^2}{\pi}} e^{-\frac{2\omega a^2}{3}}\,. 
    \label{eq:wkb_pred} 
\end{equation}
In fact, the energies can be determined for this system by solving the Ricatti equation iteratively though the Exact WKB method relatively easily to arbitrary order~\cite{iwaki2014exact,DELABAERE1997180,Delabaere:1997srq,Jentschura:2001kc,Bucciotti:2023trp,vanspaendonck2024exact}.  However, our focus here is not on precision but on the role of collective coordinates in the path integral version of the calculation.

To compute the energy splitting with the path integral, it is helpful to set up the
problem so that the leading result as $T \rightarrow \infty$ is directly
proportional to $\Delta_0$. Defining $E_n^{\pm} = E_n \mp \frac{\Delta_n}{2}$ the partition function is
\begin{equation}
  Z = \text{Tr} [ e^{- H T}] = \sum_n (e^{- E_n^+ T} + e^{- E_n^- T})\ ,
\end{equation}
and the twisted partition function is
\begin{equation}
  Z_P = \text{Tr} [P e^{- H T}] = \sum_n (e^{- E_n^+ T} - e^{- E_n^- T}) \,.
  \label{ZPEn}
\end{equation}
$Z_P$ is useful for extracting the energy since $E_n^+ = E_n^-$ to all orders in
perturbation theory; that is, the splittings are non-perturbative. 
Since $Z - Z_P$ involves only odd modes and $Z + Z_P$
involves only even modes we see that
\begin{equation}
  \lim_{T \rightarrow \infty} -\frac{1}{T} \ln \left( \frac{Z - Z_P}{Z + Z_P}
  \right) = E_0^- - E_0^+ = \Delta_0\,. \label{nosmallDelta}
\end{equation}
The above formula makes no assumption about $\Delta_0$ being small, but is useful
only if $Z$ and $Z_P$ are known non-perturbatively in $\Delta_0$. If we only know
$Z_P$ in the small $\Delta_n$ limit, then we can use
\begin{equation}
  \lim_{T \rightarrow \infty} \frac{2}{T} \Delta_n \lim_{\Delta_n \rightarrow 0}\frac{1}{\Delta_n}
  \frac{Z_P}{Z} = \lim_{T \rightarrow \infty} \frac{2}{T} \frac{\sum e^{-
  E^+_n T} (\Delta_n T)}{\sum e^{- E^+_n T}} = \Delta_0\,.
  \label{smallDelta}
\end{equation}
This formula will allow us to compute the splitting of the lowest energy states
directly with the path integral:
if we compute $Z_P$ and $Z$ in the saddle-point approximation and find at small splitting and large $T$ that $Z_P \sim Z A T$, then the energy splitting is $\Delta_0 = 2A$.

The twisted partition function can be computed with the path integral using
twisted boundary conditions:
\begin{equation}
  Z_P = \text{Tr} (P e^{- H T}) = \mathcal{N} \int_{- \infty}^{\infty} d b\ 
  \langle - b | e^{- H T} | b \rangle = \mathcal{N} \int_{- \infty}^{\infty} d
  b \int_{x \left( - \frac{T}{2} \right) = - b}^{x \left( \frac{T}{2} \right)
  = b} \mathcal{D} x\ e^{- S [x]}\,.
\end{equation}
All paths involved go from $- b$ to $b$ in time $T$ and we integrate over $b$
to take the trace. For the normalization $\mathcal{N}$, we can match to the partition function of the harmonic oscillator.

\begin{figure}
    \centering
    \includegraphics[scale=0.6]{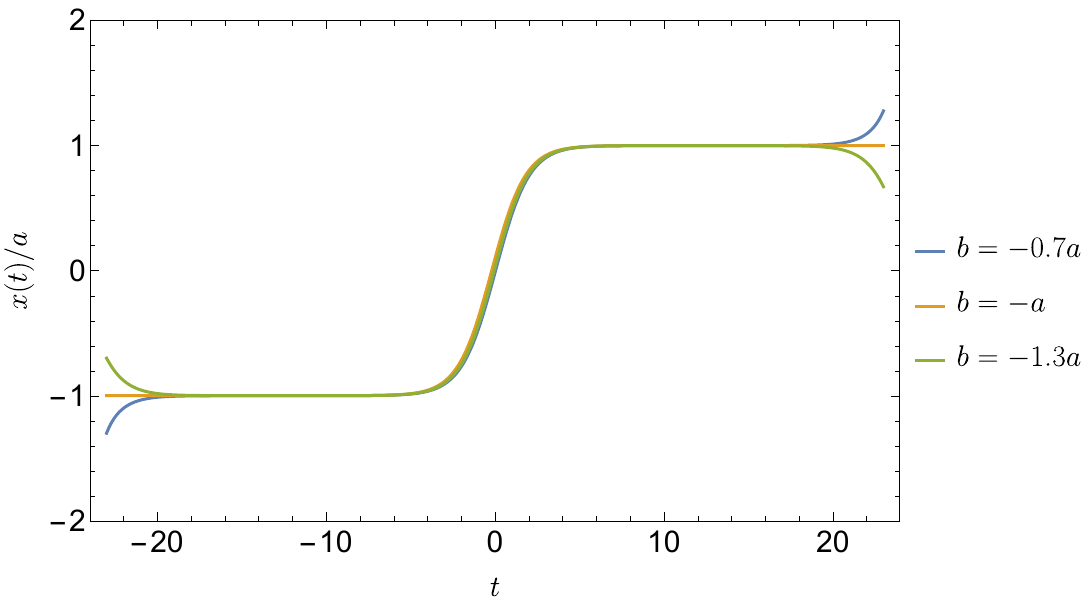}
    \caption{The twisted partition function sums over paths starting  and ending at any $b$. At finite $T$, the double well admits exact instanton solutions for any $b$. For any $b$ and large values of $T$, these solutions spend most of the time in one of the two wells. }
    \label{fig:paths}
\end{figure}
For any $b>0$, the classical path from $- b$ to $b$ for a large time $T$ is
exponentially close to the $T=\infty$ instanton
\begin{equation}
  x_I (t) = a \tanh \frac{\omega t}{2}\,.
\end{equation}
For $b<0$, the classical path from $- b$ to $b$ for a large time $T$ is closer to the anti-instanton solution $-x_I(t)$. 
Both the instanton and anti-instanton have an action given by
\begin{equation}
  S_I \equiv S [x_I] = a^2 \omega \tanh \frac{\omega T}{4} - \frac{a^2
  \omega}{3} \tanh^3 \frac{T \omega}{4} = \frac{2 a^2 \omega}{3} + \mathcal{O}
  (e^{- T \omega})\,.
\end{equation}
 In particular for $b \neq a$ these (anti-)instanton paths start at $- b$, shoot up to the top of hill of the inverted potential, stay there for almost the entire time,
roll across, and then stay at the other hill, ending at $b$ after a total time $T$. Examples (found numerically with the shooting method) that have $b$ slightly lower or larger than $a$ are shown in Fig.~\ref{fig:paths}. In particular, the part of the motion between $-b$ and $-a$ is given by a solution to the Euclidean equations of motion
\begin{align}
    \tilde{x}_b(t)= -a \coth{\frac{\omega t}{2}}\,,
\end{align}
and provides the corresponding action
\begin{align}
    S[\tilde{x}_b(t)] = \int_{a}^{b}dx\ \sqrt{2V(x)}
    =\frac{\omega}{6a}(a-b)^2(2a+b)\,.
\end{align}
Thus in the large $T$ limit, we find that the classical path from $-b$ to $b$ ($b>0$) has action 
\begin{align}
    S[x_b(t)] = S[x_I]+ 2S[\tilde{x}_b(t)]= S_I + \frac{\omega}{3a}(a-b)^2(2a+b)\,.
\end{align}
Note that when $\left| \frac{b}{a} \right| \gg 1$ then
 $S [x] \approx \frac{2 a^2 \omega}{3} +
\frac{\omega |b|^3}{3 a}$
and so the part of the action arising from the path approaching the
hills is significant.   The effect is that the contributions to the twisted partition function with $|b| \gg a$ will be small,
however we still need to include the $b$ dependence to get the numerical prefactor correct.
 
For the fluctuations, the tail regions before and after $|x| \approx a$ give exponentially small contributions, so 
we can expand paths as $x (t) \rightarrow x_I (t) + x (t)$. Then the twisted partition
function in the leading saddle point approximation becomes
\begin{align}
  Z_P &\approx 2\mathcal{N}\int_{0}^\infty d b\ e^{- S_I} e^{-2S[\tilde{x}_b(t)]} \int_{x \left( - \frac{T}{2} \right) =
  0}^{x \left( \frac{T}{2} \right) = 0} \mathcal{D} x\, \exp \left[ - \frac{1}{2} \int_{-
  \frac{T}{2}}^{\frac{T}{2}} d t\, x (t) [- \partial_t^2 + V'' (x_I)] x (t)
  \right]\\
  &\approx \mathcal{N} \sqrt{\frac{4\pi}{\omega}} e^{-S_I}\ \int_{x \left( - \frac{T}{2} \right) =
  0}^{x \left( \frac{T}{2} \right) = 0} \mathcal{D} x\, \exp \left[ - \frac{1}{2} \int_{-
  \frac{T}{2}}^{\frac{T}{2}} d t\, x (t) [- \partial_t^2 + V'' (x_I)] x (t)
  \right] 
\end{align}
where we have performed the integral over $b$ in the large $S_{I}$ limit. The factor of $2$ in front accounts for the anti-instanton solutions, which go from $-b$ to $b$ where $b<0$, and give the same contribution as the instanton solutions. Note that the instantons saturate the twisted boundary conditions and the fluctuations are Dirichlet. Next, we want to compute the path integral in the leading saddle-point approximation, carefully treating the collective coordinate and Gaussian integration over the fluctuations. 

\subsection{Normal modes at finite $T$}
Fluctuations around the instanton are described by normal modes satisfying
\begin{equation}
\label{E:PT1}
  \big[- \partial_t^2 + V'' (x_I)\big]\, x_n = \lambda_n x_n
\end{equation}
where $V'' (x_I) = \omega^2 \left( 1 - \frac{3}{2} \sech^2 \frac{\omega t}{2}
\right)$. Upon rescaling $t \rightarrow t/2$, we recognize that Eq.~\eqref{E:PT1} is equivalent to the time-independent Schr\"{o}dinger equation with a P\"{o}schl-Teller potential~\cite{Poschl:1933zz} having $\lambda = 2$. The spectrum of the differential operator in 
Eq.~\eqref{E:PT1} has
bound states and a continuum. The continuum becomes discrete at finite $T$ and
we can write a trajectory as
\begin{equation}
  x (t) = x_I (t) + a_0 x_0 (t) + a_1 x_1 (t) + \sum_{k}(a_k^+ x_k^{+} (t) + a_k^- x_k^{-} (t))
  \label{xIdw}
\end{equation}
with modes defined as follows.

The first bound state is\footnote{This function as written has eigenvalue exactly $0$, but does not satisfy the Dirichlet boundary conditions. If we impose the Dirichlet boundary conditions at $t=\pm\frac{T}{2}$, the eigenvalue gets lifted to $\lambda_0 \approx e^{-\omega T}$ (see~\cite[App. E]{Gildener:1977sm} or~\cite[Eq. 23.55b]{Muller-Kirsten:2012wla}). While an 
exponentially small eigenvalue renders finite the functional determinant in the path integral, it does not produce the correct answer for the twisted partition function because higher order terms in fluctuations in the $x_0$ direction are important. Collective coordinates allow for inclusion of fluctuations due to time translations to all orders. 
}
\begin{equation}
  x_0 (t) = \sqrt{\frac{3 \omega}{8}} \frac{1}{\cosh^2 \frac{\omega t}{2}} =
  \frac{\sqrt{6}}{2} \,\dot{x}_I (t), \quad \lambda_0 \approx 0 \,. \label{eq:inst_form}
\end{equation}
Note that $x_0 (t)$ is proportional to the
velocity of the instanton.

\begin{figure}
    \centering
\includegraphics[scale=0.7]{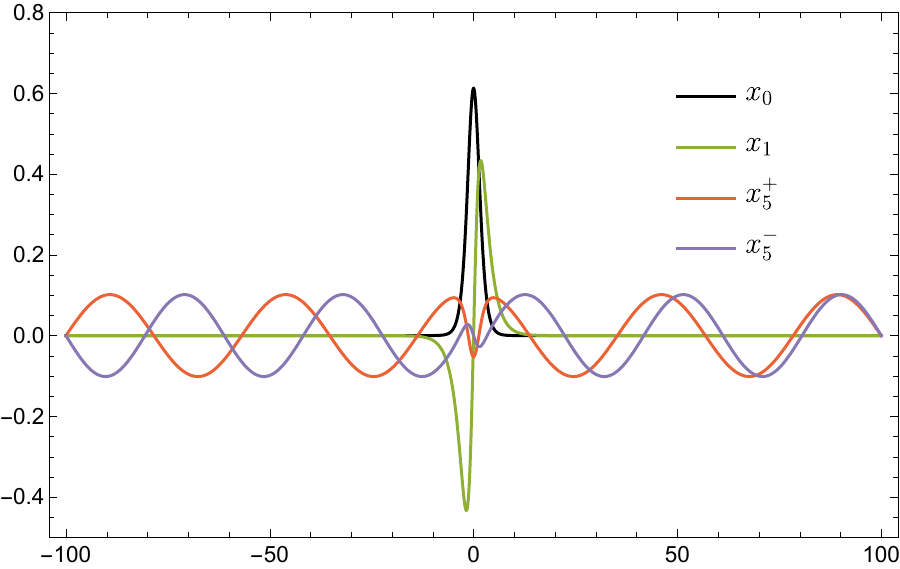}
    \caption{Normal modes of the double well at finite $T$ (plotted with $T = 200$), with Dirichlet boundary conditions at $t=\pm\frac{T}{2}$. Modes have even or odd parity. The two bound state modes $x_0$ and $x_1$ are shown, as are two of the continuum modes $x_k^\pm$ with $k=5$. These continuum modes are similar to sine functions except near $t=0$.}
    \label{fig:doublewellmodes}
\end{figure}

The second bound state has odd parity and finite
eigenvalue:
\begin{equation}
  x_1 (t) = \frac{\sqrt{3 \omega}}{2} \frac{\sinh \frac{\omega t}{2}}{\cosh^2
  \frac{\omega t}{2}}, \quad \lambda_1 = \frac{3}{4} \omega^2 \,.
\end{equation}
For $T \rightarrow \infty$ the remaining modes form a continuum. However, if we consistently maintain boundary conditions at finite $T$, these remaining modes are discrete. The eigenvalue equation is solved by
\begin{equation}
  x_k (t) = \frac{1}{4} \left( 1 + 4 \frac{k^2}{\omega^2} + 6 i \frac{k}{\omega} \tanh \frac{\omega t}{2} - 3
  \tanh^2 \frac{\omega t}{2} \right) e^{i k t}, \quad \lambda_k = \omega^2 + k^2 \,.
\end{equation}
Paths $x_k(-t)$ are also eigenstates with the same eigenvalues. 
Since the P\"{o}schl-Teller potential is symmetric, it is sensible to use parity eigenstates.  Explicitly, these modes are 
\begin{align}
  x_k^+ (t) &=N_k [x_k (t) +x_k (- t)] \\
  &= \frac{N_k}{2} \left[ \left(1 +  \frac{4k^2}{\omega^2}\right) \cos (k t) -  \frac{6k}{\omega} \sin
  (k t) \tanh \left( \frac{\omega t}{2} \right) - 3 \cos (k t) \tanh^2 \left(
  \frac{\omega t}{2} \right) \right] \nonumber
\end{align}
with $N_k = \sqrt{\frac{2\omega^4}{T(4k^4+5k^2\omega^2+\omega^4)}}$\,, and
\begin{align}
 x_k^- (t) &=-i N_k [x_k (t) -x_k (- t)]\\
 &= \frac{N_k}{2} \left[  \left(1 + \frac{4k^2}{\omega^2}\right) \sin (k t) +  \frac{6k}{\omega} \cos
  (k t) \tanh \left( \frac{\omega t}{2} \right) - 3 \sin (k t) \tanh^2 \left(
  \frac{\omega t}{2} \right) \right]\,.\nonumber
\end{align}
Note that the modes $x_k^\pm(t)$ are manifestly real.
As written, the modes are normalized at large $T$ to
$\langle  x_k^{\pm} | x_k^{\pm}\rangle =
1 + \mathcal{O} \left( \frac{1}{T} \right)$. The eigenvalues are quantized by
the boundary conditions \ $x_k^{\pm} \left( \frac{T}{2} \right) = 0$ which
leads at large $T$ to
\begin{equation}
  k^- (n) \approx \frac{2 n \pi}{T}, \quad k^+ (n) \approx \frac{(2 n - 1)
  \pi}{T}, \quad n = 1, 2, 3, ...\label{kofn} \,.
\end{equation}
As $T \rightarrow \infty$, the modes approach
\begin{equation}
  x_k^{\pm} (t) \rightarrow s_n (t) 
,
\quad \lambda_n \approx \omega^2 + \left( \frac{n \pi}{T} \right)^2, \quad n
  = 1, 2, 3,...
\end{equation}
with $s_n$ as in Eq.~\eqref{sndef}. As $T\to \infty$ the
$x_k^+$ become the even-$n$ modes and $x^-_k$ become the odd-$n$ modes.  The modes $x_k^{\pm}$ and $s_n$ differ only close to $t = 0$, as shown in Fig.~\ref{fig:doublewellmodes}. For small $t$, it is enough to know the values of the modes and their
derivatives at $t=0$:
\begin{equation}
  x_k^+ (0) = \frac{1}{\sqrt{2 T}}, \quad x_k^- (0) = 0, \quad \dot{x}_k^+ (0)
  = 0, \quad \dot{x}_k^- (0) = \frac{4\sqrt{2} n \pi}{T^{3 / 2}} \,.
\end{equation}
When $t$ is not small, the $s_n$ are excellent approximations to the $x_k^{\pm}$.

\subsection{Collective coordinates}
\noindent
Now we would like to change to collective coordinates. To do so we need to consider
the intersection number $\Ndelta$. This is the number of times a given path
$x (t)$ has vanishing projection against $x_0 (t + t_0)$ over the integration
range of $t_0$. Parameterizing $x (t)$ as in Eq.~\eqref{xIdw}, we need to find
the number of zeros of
\begin{multline}
  \big\langle x_0 (t + t_0) \big|x (t) \big\rangle = \big\langle x_0 (t + t_0) \big|x_I (t)
  \big\rangle + a_0 \big\langle x_0 (t + t_0) \big|x_0 (t) \big\rangle + a_1 \big\langle x_0 (t +
  t_0) \big|x_1 (t) \big\rangle \label{projection}
  \\[2mm] \qquad \qquad 
  + \sum_{k}\Big[a_k^{+} \big\langle x_0 (t + t_0) \big|x_k^{+} (t) \big\rangle + a_k^{-} \big\langle x_0 (t + t_0) \big|x_k^{-} (t) \big\rangle\Big]
\end{multline}
for fixed values of the coordinates $\{ a_0, a_1, a_k^{\pm} \}$. To evaluate the above
expression we note that there are three relevant time scales: the overall time
$T$, the collective coordinate time $t_0$, and the inverse frequency $\omega^{-
1}$ associated with the curvature of the potential. The two bound states
$x_0$ and $x_1$ have support of order $\omega^{- 1}$ while the continuum modes
have support over all $t$. For $t_0 \ll \omega^{- 1} \ll T$, we have
\begin{equation}
  \langle x_0 (t + t_0) |x (t) \rangle = - \sqrt{\frac{2 a^2 \omega}{3}} \,t_0 +
  a_0 - \frac{3 \pi}{16 \sqrt{2}} \,a_1 \omega t_0 + \cdots\,.
  \label{projdw}
\end{equation}
If we are permitted to drop the $a_1$ term and the additional terms thereafter, the above reduces to $t_0 \approx
 a_0 \,S^{-1/2}_I$. This fiducial trade-off between $a_0$ and $t_0$ is the change to collective coordinates one normally
anticipates: motion along the direction of the zero mode $x_0$ is exchanged
for motion along $t_0$. However, the relationship between $a_0$ and $t_0$ only holds for $t_0 \ll
\omega^{- 1} \ll T$. For larger $t_0$ the other terms in Eq.~\eqref{projdw}
are important. 

In the regime $\omega^{- 1} \ll t_0 \ll T$, the detailed structure of $x_0 (t
+ t_0)$ of order $\omega^{- 1}$ cannot be resolved. In fact, as $\omega t_0 
\rightarrow \infty$, the mode $x_0(t-t_0) \sim \frac{1}{\cosh^2 \frac{\omega (t-t_0)}{2}}$
approximates a $\delta$-function:
\begin{equation}
  x_0 (t - t_0) \approx \sqrt{\frac{6}{\omega}}\, \delta (t -t_0)\,,\quad 
  \omega^{- 1} \ll  t_0 \ll T\,,
\end{equation}
so that
\begin{equation}
  \big\langle x_0 (t - t_0) \big|x (t) \big\rangle \approx \sqrt{\frac{6}{\omega}} \int d t\,
  \delta (t - t_0) x (t) = \sqrt{\frac{6}{\omega}}\, x (t_0) \,.
  \label{origincross}
\end{equation}
The equation above leads to a nice interpretation of $N_{x_0}[x]$: it is the number of times
$x (t_0) = 0$, i.e.~the number of times a given path $x (t)$ crosses $x = 0$
on its way from $-b$ to $b$. Similarly, in the same limit the Jacobian in Eq.~\eqref{Zcollmain} $\mathcal{J} = \langle \dot{x}_0 (t) |x (t) \rangle \approx
- \dot{x} (0)$ becomes the velocity of the path at $t_0$, i.e.~as the path crosses the origin.

\subsection{Energy splitting \label{sec:splitting}}

Now let us use the collective coordinates to compute the energy splitting. To
do so, it is easiest to parameterize paths by
\begin{equation}
  x (t) = x_I (t + t_0) + \alpha_1 x_1 (t + t_0) + \sum_k \Big[\alpha_k^{+} x_k^{+} (t +
  t_0) + \alpha_k^{-} x_k^{-} (t +
  t_0)\Big] \label{alphadw}
\end{equation}
and to use Eq. \eqref{Zform1}. The Jacobian evaluates to
\begin{align}
  \mathcal{J} &= \big\langle \dot{x}_0 (t) \big|x_I (t) + \alpha_1 x_1 (t) +
  \sum_{k}\Big[\alpha_k^{+} x_k^{+} (t) + \alpha_k^{-} x_k^{-} (t)\Big] \,\big\rangle
  \\*
 &\approx 
- \sqrt{\frac{2 \omega a^2}{3}} - \frac{3 \pi \omega}{16 \sqrt{2}}
  \alpha_1 + \sum_{k}\alpha_k^+ \frac{4\sqrt{2} n \pi}{T^{3 / 2}} \label{Jform3}
\end{align}
where $n$ refers to the approximate wavenumber from Eq.~\eqref{kofn}.

The action of a path $x(t)$ parameterized in collective coordinates is
\begin{equation}
  S [x] = \frac{1}{2} \alpha_1^2 \lambda_1 + \frac{1}{2} \sum_{k}\Big[(\alpha_k^{+})^2 \lambda_k^{+}+ (\alpha_k^{-})^2 \lambda_k^{-}\Big]   \,.
\end{equation}
Since $\lambda_1 = \frac{3}{4} \omega^2$ and $\lambda_k \approx \omega^2 +
\left( \frac{n \pi}{T} \right)^2$, the above action implies that the path integral is dominated by the regime $\alpha_1 \lesssim \omega^{- 1}$ and $\alpha_k^\pm  \lesssim \omega^{- 1}$ (and even for $n > T
\omega$ the scaling is $\alpha^{\pm}_k \lesssim \frac{T}{n} \lesssim \omega^{-1}$ as well). Thus, for purposes of computing the energy splitting where we are
interested in the limit of large $a^2 \omega$ and large $T$, the Jacobian
factor in Eq.~\eqref{Jform3} reduces to
\begin{equation}
   \mathcal{J}  \approx -\sqrt{\frac{2 \omega a^2}{3}} = -\sqrt{S_I} \,.
  \label{jacobian} 
\end{equation}
This is the result usually found in the literature.

In collective coordinate, the twisted partition function is then
\begin{equation}
  Z_P =  \mathcal{N} \sqrt{\frac{4\pi}{\omega}} \sqrt{S_I} \,e^{- S_I} \int_{-
  \frac{T}{2}}^{\frac{T}{2}} \frac{d t_0}{\sqrt{2\pi}} \int_{- \infty}^{\infty} \frac{d \alpha_1}{\sqrt{2\pi}}\, \prod_{k}\int_{-\infty}^{\infty}\frac{d
  \alpha_k^{\pm}}{\sqrt{2\pi}}\ e^{-\frac{1}{2} \lambda_1 \alpha_1^2 -\frac{1}{2} \lambda_k (\alpha_k^{\pm})^2}
  \frac{1}{\Nxzero}
\end{equation}
where, as discussed, $\Nxzero$ counts the number of times a path $x
(t)$ crosses $x = 0$.   Although $\Nxzero$ is an incredibly complicated
functional of paths, it can be simplified for the calculation of the twisted partition function in the large $T$ limit.

We reiterate that according to Eq.~\eqref{origincross}, $\Nxzero$ counts the number of times the path crosses the origin. 
Since all paths integrated over
in the twisted partition function go from $-b$ to $b$ they must all cross the
origin at least once -- this is essential, otherwise the change to collective
coordinate $t_0$ would not be valid. 
Because the potential $V (x)$ is largest at $x = 0$, larger values of
$\Nxzero$ will give exponentially suppressed contributions to the path integral compared with the contributions at $\Nxzero=1$. Consequently $\Nxzero \geqslant 1$ and
paths with $\Nxzero > 1$ are negligible. Thus we can approximate $\Nxzero
= 1$. So we have
\begin{align}
  Z_P &\approx \mathcal{N} \sqrt{\frac{4\pi}{\omega}} T  \sqrt{\frac{S_I}{2\pi}} \,e^{- S_I} \int_{-\infty}^{\infty}\frac{d
  \alpha_1}{\sqrt{2\pi}} \prod_{k} \int_{-\infty}^{\infty}\frac{d
  \alpha_k^{\pm}}{\sqrt{2\pi}}\ e^{-\frac{1}{2} \lambda_1 \alpha_1^2 - \frac{1}{2}\lambda_k (\alpha_k^{\pm})^2}
  \\*
  &= \mathcal{N} \sqrt{\frac{4\pi}{\omega}} T \sqrt{\frac{S_I}{2\pi}} \,e^{- S_I} \sqrt{\frac{1}{\lambda_1}}
  \prod_k \sqrt{\frac{1}{\lambda_k^{\pm}}} \\
  &= \mathcal{N} \sqrt{\frac{4\pi}{\omega}} T \sqrt{\frac{S_I}{2\pi}} \,e^{- S_I} \left[\text{det}^{\prime}(-\partial_t^2 +V''(x_I))\right]^{-1/2} \label{eq:dw_with_norm} \,.
\end{align}
As such we have reduced our calculation to the textbook treatment.

We can also understand the $\Nxzero \approx 1$ replacement from a different perspective, which allows a more straightforward generalization to the QFT case in Section~\ref{sec:qft}. If we solve the condition $\langle x_0(t+t_0) | x(t)\rangle=0$ for $t_0$ using Eq.~\eqref{projdw} and expand at large $S_I$, we find
\begin{equation}
    t_0 = \frac{a_0}{\sqrt{S_I}} - \frac{3\pi \omega}{16\sqrt{2} S_I} \,a_0\,a_1 + \cdots \label{t0sol} \,.
\end{equation} 
Here $a_n \lesssim \omega^{-1}$ just as $\alpha_n \lesssim \omega^{-1}$. However, the mode with zero eigenvalue with coordinate $a_0$ is unsuppressed. Considering only the zero mode, the condition determining $t_0$ is
\begin{align}
0 &=    \langle x_0 (t + t_0) |x_I (t)
  \rangle + a_0 \langle x_0 (t + t_0) |x_0 (t) \rangle 
  \label{E:firstline1}
  \\
  & = \frac{3}{\omega}\sqrt{ S_I}\,\frac{(\omega t_0 -\sinh (\omega t_0) )}{\cosh \omega t_0 - 1} +a_0\,\frac{3  (-2 +\omega t_0 \coth \frac{\omega t_0}{2} )}{\cosh \omega t_0 - 1}   \,.
\end{align}
The first function on the second line is a monotonically decreasing odd function of $t_0$ and the second function is a positive monotonically decreasing function of $|t_0|$ with a maximum at $t_0 = 0$; these conditions imply that there is exactly one solution to~\eqref{E:firstline1} for a given $a_0$. 
There are more solutions for $t_0$ if we turn on some of the other $a_n$'s, analogous to the non-linear terms in   Eq.~\eqref{t0sol}.  Such additional solutions for $t_0$ require large values of $a_n \gtrsim \sqrt{S_I}$. However, when $a_n \gtrsim \omega^{-1}$, the path integral becomes exponentially suppressed. Thus any path with $\Nxzero > 1$ is  exponentially suppressed by the contributions of large $a_n$ and so we can simply set $\Nxzero=1$.

The above argument can be stated more physically. We expand around a particular instanton which crosses the origin at $t_0=0$. We can shift the path using the zero mode, changing the time the path crosses the origin. But to get an additional crossing, the quantum corrections must be enormous: as large as the classical contribution. Such large corrections are exponentially suppressed.

There is a close analogy between this calculation, where the energy splitting is given by paths that cross $x=0$ exactly once, and the calculation of the tunneling rate in an asymmetric double well. In the latter, one way to derive the formula for the decay rate is to restrict to paths that cross the origin only once~\cite{Andreassen:2016cff,Andreassen:2016cvx}. This restriction can be understood physically: if the paths do not cross the origin at all, they do not mediate transitions from the false vacuum to the true one; to find a well-defined decay rate one must be in a regime where the back-scattering off the far end of the potential is irrelevant, which can be enforced by demanding at most one crossing. Thus the decay rate is determined by paths that cross the origin exactly once, akin to the energy splitting in the symmetric double well.

For completeness, we finish the energy-splitting calculation. The product over continuum
eigenvalues is nearly the same as for a simple harmonic oscillator. For the
harmonic oscillator, one expands around the classical solution $x_{\text{cl}}(t)=b \cosh(\omega t)/\cosh(\omega T/2)$ where the fluctuations have eigenvalues $\lambda_n^{\text{SHO}} = \omega^2 + ( {n
\pi}/{T} )^2$, and so the partition function is
\begin{align}
  Z_{\text{SHO}} &= \mathcal{N} \int_{-\infty}^{\infty}db\ e^{-\omega b^2}\,  \prod_{n}\int_{-\infty}^{\infty} \frac{da_{n}}{\sqrt{2\pi}}\, e^{-\frac{1}{2}\lambda_{n}a_n^2} \\
  &=
  \mathcal{N} \sqrt{\frac{\pi}{\omega}}\prod_n \sqrt{\frac{1}{\lambda_n^{\text{SHO}}} }=\mathcal{N} \sqrt{\frac{\pi}{\omega}}\ \text{det}^{-1/2}(-\partial_t^2+\omega^2)
\end{align}which defines for us the normalization $\mathcal{N}$ in terms of $Z_{\text{SHO}}$.  For the double well, we can then plug $\mathcal{N}$ into Eq.~\eqref{eq:dw_with_norm}, obtaining
\begin{align}
    Z_{P}=Z_{\text{SHO}}\ 2T\sqrt{\frac{S_I}{2\pi}} \,e^{-S_{I}} \left(\frac{\det(-\partial_t^2 +\omega^2)}{\text{det}^{\prime}(-\partial_t^2 +V^{\prime\prime}(x_{I}))}\right)^{1/2} \,.
\end{align}
The ratio of determinants can be computed using the Gelfand-Yanglom method \cite{10.1063/1.1703636,Dunne:2007rt} and yields $12\omega^2$.  Using Eq.~\eqref{smallDelta}, we find
\begin{equation}
  \Delta_0 = 2\sqrt{\frac{6 S_I}{\pi}} \,e^{- S_I} \omega 
\end{equation}
Recalling $S_I = \frac{2 \omega a^2}{3}$, 
the above exactly matches the WKB result in Eq.~\eqref{eq:wkb_pred}. We have used that $Z \approx 2 Z_{\text{SHO}}$ in the small-energy-splitting limit.

\subsection{Free twisted partition function}

In the interacting theory, integration over the collective coordinate $t_0$
gave a factor of $T$ in the twisted partition function, as expected for the
computation of the energy splitting. The free theory is also time-translation
invariant, and so using collective coordinates would seem to give a factor of $T$.
However, the free twisted partition function is just $Z_P = \frac{1}{2}$ (as
we review shortly). So what is the fate of the putative factor of $T$? To find out, we need to properly
compute the twisted partition function with action $S_0 [x] = \int d t\,
\frac{1}{2} \dot{x}^2$ using collective coordinates.

\subsubsection*{Expected results}

First we recall the expected result. The twisted partition function in the free theory can be
computed as the zero frequency limit of the twisted partition function for the
harmonic oscillator, namely
\begin{equation}
  Z_P^0 = \text{Tr} (P e^{- H T}) = \lim_{\omega \rightarrow 0}\, \sum_{n=0}^{\infty}e^{- \omega
  \left( n + \frac{1}{2} \right) T} (- 1)^n = \frac{1}{2} \,.
\end{equation}
Alternatively, we can recall that the position-space propagator in quantum
mechanics is
\begin{equation}
  D (a, b) = \langle b | e^{- H T} | a \rangle = \int_{-\infty}^{\infty}\frac{dp}{2\pi}\langle b|e^{- \frac{1}{2}
  p^2 T} |p \rangle \langle p|a \rangle = \int_{- \infty}^{\infty} \frac{d
  p}{2 \pi} \,e^{i p (b - a)} e^{- \frac{p^2}{2} T} = \sqrt{\frac{1}{2 \pi T}}
  \,e^{- \frac{(b - a)^2}{2 T}} \,.
\end{equation}
So the twisted partition function is
\begin{equation}
  Z_P^0 = \int_{- \infty}^{\infty} d b\ D (b, - b) = \int_{- \infty}^{\infty} d
  b\, \sqrt{\frac{1}{2 \pi T}} \,e^{- \frac{2 b^2}{T}} = \frac{1}{2} \,.
  \label{twistfree}
\end{equation}

In the path integral approach, we compute
\begin{equation}
  Z_P^0 = \mathcal{N} \int_{- \infty}^{\infty} d b \int_{x \left( -
  \frac{T}{2} \right) = - b}^{x \left( \frac{T}{2} \right) = b} \mathcal{D} x\ 
  e^{- S [x]} \,.
\end{equation}
Here the classical path going from $- b$ to $b$ in time $T$ is $x_c (t) =
\frac{2 t}{T} b$ with action $S_0 [x_c] = \frac{2 b^2}{T}$, as expected from
the exponent of  $D (b, - b)$. A general path is then parameterized by
\begin{equation}
  x (t) = x_c (t) + c_n s_n (t), \quad n = 1, 2, 3, \ldots\,,\label{cnbasis}
\end{equation}
with $s_n (t)$ as in Eq. \eqref{sndef}. These $s_n(t)$ modes satisfy $\partial_t^2 s_n = -
\lambda_n s_n$ with $\lambda_n = \left( \frac{n \pi}{T} \right)^2$. Then
\begin{equation}
  Z_P^0 = \mathcal{N} \int_{- \infty}^{\infty} d b\ e^{- \frac{2 b^2}{T}}
  \prod_{n=1}^{\infty}\int_{- \infty}^{\infty} \frac{d c_n}{\sqrt{2\pi}}\ e^{- \frac{1}{2} \lambda_n c_n^2} =
  \mathcal{N} \frac{\sqrt{2 \pi T}}{2} \prod_n \frac{T}{n  \pi}
\end{equation}
so that to have $Z_P^0 = \frac{1}{2}$ we need
\begin{equation}
  \mathcal{N} = \frac{1}{\sqrt{2 \pi T}} \left[ \prod_n 
  \frac{T}{n  \pi} \right]^{- 1} \,.
\end{equation}

\subsubsection*{Collective coordinates}
Now we attempt to compute $Z_P^0$ using collective coordinates in
the double-well basis. Noting that the formula in Eq.~\eqref{Zcollmain} is
basis-independent, the only feature from the double well basis we carry over is that the function $X (t) = x_0 (t)$ is the one used to project out the collective
coordinate. Indeed, if we had tried to use the parametrization in Eq.~\eqref{xIdw} for the paths in the free theory case we would have run into
problems because $x_I (t)$ is not a saddle point of the free action (in fact, it has  much larger action than the classical path; for paths from $-a$ to $a$, $S_0 [x_I] =
\frac{a^2 \omega}{3} \gg S_0 [x_c] = \frac{2 a^2}{T} )$.  Since $x_I(t)$ is not a saddle point, in attempting to use Eq.~\eqref{xIdw} in the path integral, we would find linear fluctuation terms which would require completing the square in the action,
effectively reducing the computation to expanding around the classical path.
The result would be the same as 
using the parameterization in Eq.~\eqref{cnbasis} to
begin with.

Marching forward, we want to evaluate
\begin{equation}
  Z_P^{\text{coll}, 0} = \mathcal{N} \int_{- \infty}^{\infty} d b\ e^{- \frac{2
  b^2}{T}} \int_{- \frac{T}{2}}^{\frac{T}{2}} d t_0\ \prod_{n}\int_{-\infty}^{\infty} \frac{d c_n}{\sqrt{2\pi}}\ e^{- \frac{1}{2}
  \lambda_n c_n^2} \frac{1}{\Nxzero} \,\delta \Big[\big\langle x_0 (t) \big|x (t)
  \big\rangle \Big] \, \Big| \big\langle x_0 (t) \big| \dot{x}  (t) \big\rangle \Big| \,.
\end{equation}
As discussed in the double-well case, $\delta [\langle x_0 (t) |x (t) \rangle]$
forces the path to cross $x = 0$ at $t = 0$ and $\Nxzero$ counts the
number of times a given path crosses $x = 0$.  To compute the argument of the $\delta$ function we need the projections
\begin{align}
  \langle x_0 (t) |s_n (t) \rangle &= \sqrt{\frac{2}{T}} \int_{-
  \infty}^{\infty} d t\ x_0 (t) \sin \frac{n \pi \left( t + \frac{T}{2}
  \right)}{T} = \sqrt{\frac{12}{T\omega}} \sin \frac{n \pi}{2} \,Q_n
  \label{moreprecise}
  \\
  &= (- 1)^{\frac{n - 1}{2}} \sqrt{\frac{12}{T\omega}} \,\delta_{n,\,\text{odd}} \,Q_n
\end{align}
which we calculate in the $T\rightarrow \infty$ limit, where 
\begin{equation}
  Q_n = \frac{\frac{n \pi^2}{T\omega}}{\sinh \frac{n \pi^2}{T\omega}} \approx 1 \,.
\end{equation}
Thus the $\delta$ function involves only the odd modes, and sets $c_1 - c_3 +
c_4 - c_5 + \cdots = 0$ up to corrections in $\frac{1}{T}$. The Jacobian is
\begin{equation}
  \langle x_0 (t) | \dot{x} (t) \rangle = \langle x_0 (t) | \dot{x_c} (t)
  \rangle + \sum_n c_n \langle x_0 (t) | \dot{s}_n (t) \rangle = \frac{2b}{T}\sqrt{\frac{6}{\omega}} + \sum_n (- 1)^{\frac{n}{2}} \sqrt{\frac{12}{\omega T}}\sqrt{\lambda_n} \,c_n\,
  Q_n\,\delta_{n,\,\text{even}}
\end{equation}
which involves only the even modes.
Since the $\delta$ function only involves the odd modes and the Jacobian
only involves the even modes, the integration over the even and odd modes
factorizes.

With the factor of $\Nxzero$, even though the path integral factorizes, the integrals are challenging. To show that $\Nxzero$ is necessary, it suffices to ignore it by setting it to 1 and show that the wrong answer results. 
 Let us do so and behold the consequences of such ignorance. 

The formulas
\begin{equation}
  \int d^N c\ e^{- \pi \vec{c} \cdot A \cdot \vec{c}} | \vec{w} \cdot \vec{c} |
  = \frac{1}{\pi}  \frac{\sqrt{\vec{w} \cdot A_{}^{- 1} \cdot
  \vec{w}}}{\sqrt{\det (A)}}
\end{equation}
and
\begin{equation}
  \int d^N c \ e^{- \pi \vec{c}^{\,2}} \delta (\vec{w} \cdot
  \vec{c}) = \frac{1}{\sqrt{\vec{w}^2}}
\end{equation}
are helpful.  For the odd modes we find
\begin{equation}
  I_{\text{odd}} = \prod_{n\,\text{odd}} \int \frac{d c_n}{\sqrt{2\pi}} \,e^{- \frac{1}{2} c_n^2
  \lambda_n} \delta \left( c_n \sqrt{\frac{12}{\omega T}} \sin \frac{n \pi}{2} Q_n
  \right) = \left( \prod_{n\,\text{odd}} \sqrt{\frac{1}{\lambda_n}} \right)
  \sqrt{\frac{\omega}{3 \pi T}}
\end{equation}
and for the even modes we have
\begin{align}
  I_{\text{even}} &= \int d b\ e^{- \frac{2 b^2}{T}} \prod_{n \, \text{even}} \int \frac{d
  c_n}{\sqrt{2\pi}}\, e^{- \frac{1}{2} \lambda_n c_n^2} \left| \frac{2b}{T}\sqrt{\frac{6}{\omega}} + \sum_n(- 1)^{\frac{n}{2}} \sqrt{\frac{12}{\omega T}} \sqrt{\lambda_n} \,c_n\,
  Q_n\,\delta_{n, \text{even}} \right|
\\
&= \left( \prod_{n\, \text{even}} \sqrt{\frac{1}{\lambda_n}} \right) \left(
  \sqrt{T} + \frac{3}{\omega \sqrt{T}} + \cdots \right) \,.
\end{align}
So all together we obtain
\begin{align}
  Z_P^{\text{coll}, 0} &= \frac{1}{\sqrt{2 \pi T}} \left[ \prod_n \sqrt{\frac{1}{\lambda_n}} \right]^{- 1} \left( \prod_{n\,  \text{odd}} \sqrt{\frac{1}{\lambda_n}} \right) \sqrt{\frac{\omega}{3 \pi T}} \left( \prod_{n\, 
  \text{even}} \sqrt{\frac{1}{\lambda_n}} \right) \sqrt{T} \int_{-T/2}^{T/2} d t_0
  \\
  &= \sqrt{\frac{\omega T}{6 \pi^2}} \,.
  \label{E:notright1}
\end{align}
This does not agree with the correct answer $Z_P^0 = \frac{1}{2}$, as in Eq.~\eqref{twistfree}. Indeed Eq.~\eqref{E:notright1} is too large by a factor of $\sqrt{\omega T}$. Thus we confirm that using collective coordinates without the factor of $\Nxzero$ is incorrect. 

To see that the right answer does emerge if $\Nxzero$ is included is challenging because of the extreme complexity of $\Nxzero$. However, incorporating $\Nxzero$ is essentially the same challenging problem discussed in Section~\ref{sec:free}. It does not seem that working in the double-well basis rather than a Fourier basis provides any additional insight, so we simply refer the reader back to Section~\ref{sec:free} and move on to quantum field theory.

\section{Quantum field theory \label{sec:qft}}
\noindent
For our final example, we consider the $\lambda \phi^4$ quantum field theory
in 4 Euclidean dimensions.  The Lagrangian density is $\mathcal{L} [\phi] =
\frac{1}{2} \partial_{\mu} \phi \partial^{\mu} \phi + V (\phi)$ with $V (\phi)
= \frac{1}{4} \lambda \phi^4$. For $\lambda < 0$ this theory has a
5-parameter family of Fubini-Lipatov instanton saddle points~\cite{Fubini:1976jm, Lipatov:1976ny} called ``bounces'' given by
\begin{equation}
  \phi_b^{R, \xi} (x) = \sqrt{\frac{8}{- \lambda}} \frac{R}{R^2 + (x^{\mu} +
  \xi^{\mu})^2}\,,
\end{equation}
with action
\begin{equation}
  S_b = S [\phi_b] = - \frac{8 \pi^2}{3 \lambda} \,.
\end{equation}
We will generally pick a fiducial member of this family to expand around and
denote it by $\phi_b (r)$ with $R$ fixed and $\xi = 0$.  Here $r =
\sqrt{\sum_{\mu} x_{\mu}^2}$ is the Euclidean radial coordinate.

With $\lambda < 0$ the vacuum at $\phi = 0$ is unstable. The corresponding decay rate can be
computed perturbatively as
\begin{equation}
\Gamma = \lim_{T \rightarrow \infty} \frac{1}{T} \text{Im} \frac{\int_{C_b}
  \mathcal{D} \phi\, e^{- S [\phi_b] + \frac{1}{2} S'' [\phi_b] \phi^2 +
  \cdots}}{\int_{C_{\text{FV}}} \mathcal{D} \phi\, e^{- \frac{1}{2} \int d^4 x
  (\partial_{\mu} \phi)^2}} \label{FVdecay}
\end{equation}
where $C_b$ is a path of steepest descent passing though the bounce $\phi_b$
and $C_{\text{FV}}$ is the expansion around the false vacuum $\phi = 0$. A
derivation of this formula can be found in \cite{Andreassen:2017rzq}. The numerator of this
expression is expected to have a factor of $T \mathcal{V}$ from the translation
invariance of the action (the $\xi^{\mu}$ moduli). Such a factor makes $\Gamma
\propto \mathcal{V}$, i.e.~the decay is a rate per unit volume, as expected in a
translationally invariant theory. The denominator, which comprises an expansion around the
free theory, is not expected to have a factor of $T \mathcal{V}$. The goal of this section is to understand why the
numerator is proportional to $T \mathcal{V}$ but the denominator is not, despite the
fact that both actions are translation and scale invariant.

\subsection{Basis and cross check}
\noindent
Remarkably, the eigenstates for expanding around the bounce are known in
closed form~\cite{Andreassen:2017rzq}. Even more remarkably, these same eigenstates also diagonalize
fluctuations around the false vacuum. The modes are 
\begin{equation}
  \hat{\phi}_{n s \ell m} (r, \alpha, \theta, \phi) = 
  \frac{\sqrt{2 n + 3}}{2 \pi} 
  \frac{1}{r} P_{n + 1}^{- s - 1} \left( \frac{R^2 - r^2}{R^2 + r^2}
  \right)\sqrt{\frac{\pi}{2}} \,Y^{s \ell m} (\alpha, \theta, \phi)
  \label{normalmodes}
\end{equation}
where $P_{\ell}^m (x)$ are associated Legendre polynomials and $Y^{s \ell m}
(\alpha, \theta, \phi)$ are 3D spherical harmonics (which are also Legendre
polynomials). We will use the ``box'' inner product
\begin{equation}
  \langle \phi_i | \phi_j \rangle = - \int d^4 x\, \phi_i (x) \Box \phi_j (x)
\end{equation}
under which the modes are orthonormal, namely
\begin{equation}
  \langle \hat{\phi}_{n s \ell m} | \hat{\phi}_{n' s' \ell' m'} \rangle = -
  \int d^4 x\, \hat{\phi}_{n s \ell m} \Box \hat{\phi}_{n' s' \ell' m'} =
  \delta_{n n'} \delta_{s s'} \delta_{\ell \ell'} \delta_{m m'} \,.
\end{equation}
The modes satisfy 
\begin{equation}
  \Box\, \hat{\phi}_{n s \ell m} = \lambda_n V'' [\phi_b] \hat{\phi}_{n s \ell
  m}, \quad \lambda_n = \frac{(n + 1) (n + 2)}{6} = \frac{1}{3}, 1, 2,
  \frac{10}{3}, ...\,,
\end{equation}
and there are $d_n = \frac{(n + 1) (n + 2) (2 n + 3)}{6} = 1, 5, 14, ...$ modes with eigenvalue $\lambda_n$.

The $n = 0$ mode with multiplicity $d_0 = 1$ is proportional to the bounce 
\begin{equation}
  \phi_{0000} = \frac{\sqrt{3}}{2 \pi} \frac{R}{r^2 + R^2} = \frac{1}{2 \pi}
  \sqrt{\frac{- 3 \lambda}{8}} \,\phi_b = \frac{1}{2 \sqrt{S_b}} \,\phi_b \,.
\end{equation}
Normalizing with respect to our box inner product, the mode is independent of $\lambda$. Of the five
$n = 1$ modes the one with $s = 0$ is the dilatation mode associated with
changes in $R$, namely
\begin{equation}
\phi_d \equiv   \phi_{1000} = \frac{\sqrt{5} R}{2 \pi} \frac{R^2 - r^2}{(R^2 + r^2)^2} =
  -\sqrt{\frac{5}{12S_b}}R \, \partial_R \phi_b \label{dilmode} \,.
\end{equation}
The other four $n = 1$ modes, with $s = 1$, correspond to translations. They are linear combinations of the $\partial_{\mu} \phi_b$.

Field configurations around the bounce can be parameterized as
\begin{equation}
  \phi (x) = \phi_b (x) + \sum_{n} c_n \hat{\phi}_n (x)
\end{equation}
where we are using $n$ to abbreviate all the indices. The quadratic action
around the bounce is 
\begin{align}
  S [\phi] &\approx S[\phi_b]+ \sum_{n,\, m}\frac{1}{2}\int d^4 x \,  c_n \hat{\phi}_n
  [-\Box + V'' (\phi_b)]  c_m \hat{\phi}_m\\*
  &= S_b + \sum_{n}\frac{1}{2} c_n^2 \cdot\left( \frac{\lambda_n - 1}{\lambda_n} \right) \\*
  &=
  S_b + \frac{1}{2} c_0^2 \cdot (- 2) + \frac{1}{2} c_1^2\cdot (0) + \frac{1}{2} c_2^2 \cdot
  \left( \frac{1}{2} \right) + \cdots \,. \label{eq:squadbounce}
\end{align}
The unusual factor $(\lambda_n - 1)/\lambda_n$ is a result of the box inner product. If we
had used the ``$V$'' inner product of Ref.~\cite{Andreassen:2017rzq}, namely 
$\langle \phi_i | \phi_j \rangle_V = -
\int d^4 x\, V'' [\phi_b] \phi_i \phi_j$, we would have found the action $S = S_b +
\frac{1}{2} c_n^2 (\lambda_n - 1) + \cdots$. For present purposes, the box
inner product is preferable because it is well-defined even for $V = 0$ so we can
use it in the free theory.
Importantly, there is one mode ($n = 0$) with a negative eigenvalue and five
modes ($n = 1$) with zero eigenvalue in either inner product.

Writing $\phi (x) = \sum_{n}b_n \hat{\phi} _{n}(x)$, the free action in these coordinates is
\begin{equation}
  S [0] + \frac{1}{2}\int d^{4}x\, S'' [0] \phi(x)^2 = \sum_{n,\,m}\frac{1}{2}\int d^4 x (b_n \hat{\phi}_n) [-\Box] (b_m
  \hat{\phi}_m) = \sum_{n}\frac{1}{2} b_n^2 \,.
\end{equation}
The reason that the same basis diagonalizes the expansion around the bounce and
around the free theory is that the bounce is itself one of the the normal
modes (the $n = 0$ one). Thus we can move from the bounce to the false vacuum by varying a single coordinate.

For a cross check, let us compute the propagator in the free theory. We
can write the propagator in terms of the path integral as
\begin{align}
  D (x, 0) &= \frac{\int \mathcal{D} \phi\, e^{- S [\phi]} \phi (x) \phi
  (0)}{\int \mathcal{D} \phi e^{- S [\phi]}} = \frac{\int \prod_{n}d b_n\, e^{-
  \frac{1}{2} b_n^2} \sum_{k,j}[b_k \phi_k (x)] [b_j \phi_j (0)]}{\int \prod_{n}d c_n\, e^{-
  \frac{1}{2} c_n^2}}
\\*
  &= \sum_n \phi_n (x) \phi_n (0)\\*
  &= \frac{1}{4 \pi^2 R r} \sum_n (2 n + 3) P_{n + 1}^{- 1}\!\left( \frac{R^2 -
  r^2}{R^2 + r^2} \right) \,.
\end{align}
Using the integral representation
\begin{equation}
  P_{n + 1}^{- 1} (x) = \frac{1}{i \pi (n + 1)} \int_0^{\pi} d t \cos (t)
  \left( \sqrt{x^2 - 1} \cos t + x \right)^{n + 1}
\end{equation}
we can perform the sum, leading to
\begin{equation}
  D (x, 0) = \frac{1}{4 \pi^2 r^2}
\end{equation}
which is the correct answer and as expected is independent of $R$.

\subsection{Collective coordinates for dilatations \label{sec:dilatations}}
To go to collective coordinates, we consider first the $R$ coordinate
associated with dilatations. The $\lambda \phi^4$ theory is classically scale invariant,
including if we expand around the bounce or the false vacuum. Ultimately, the
divergence associated with an integration over $R$ will be regulated by the
scale anomaly, which shows up at higher orders
in the saddle point
approximation. We are not interested in this regulation here. Rather, we are
interested in showing why a divergent integral over the collective coordinate
$R$ should show up in the expansion around the bounce but not around the free theory.

Let us denote the dilatation mode as $\phi_d \equiv \phi_{1000}$. This is the
mode we denoted by $X (t)=x_0(t)$ in the quantum mechanical case, and is the mode we would like to
remove. While $R$ is already incorporated into our basis expansion, let us make the $R$-dependence explicit by fixing $R = \Rb$ and writing
\begin{equation}
  \phi (x) = \phi_b (\Rb, x) + c_d \phi_d (\Rb, x) + \sum_{n\geq 1}c_n \phi_n (\Rb, x) \,.
  \label{R0form}
\end{equation}
Then the collective coordinate parametrization where $R$ is a coordinate and
$\phi_d$ is removed is
\begin{equation}
  \phi (x) = \phi_b (R, x) + \sum_{n\geq 1} \alpha_n \phi_n (R, x) \,.
\end{equation}
In these collective coordinates, the Jacobian factor from Eq.~\eqref{Zform1} is 
\begin{equation}
  \mathcal{J} = \langle \phi_d | \partial_R \phi \rangle = -
  \sqrt{\frac{3}{5}} \frac{1}{R} \left( 2 \sqrt{S_b} + \alpha_0 \right) +
  \frac{8}{\sqrt{35}} \frac{1}{R} \alpha_2
\end{equation}
where $\alpha_2$ is the coefficient of $\phi_{2000}$. Only the radial modes
with $\ell = 0$ (and only two of them, with $n=0,2$) contribute. Accordingly the expansion of the path integral
around the bounce is
\begin{equation}
\label{E:tostudy1}
  e^{- S_b} \int \frac{d R}{R} \int \prod_{n}d\alpha_{n}\, e^{- \frac{1}{2} \alpha_n^2
  \frac{\lambda_n - 1}{\lambda_n}} \delta (\alpha_1) \left| \sqrt{\frac{3}{5}}
  \left( 2 \sqrt{S_b} + \alpha_0 \right) - \frac{8}{\sqrt{35}} \alpha_2
  \right| \frac{1}{\Ndeltaphi}
\end{equation}
where $\Ndeltaphi$ counts the number of solutions to $\langle \phi_d(R,x) | \phi(\Rb,x) \rangle=0$. For the free theory, the path integral in the collective coordinate basis is
\begin{equation}
\label{E:tostudy2}
  \int \frac{d R}{R} \int \prod_{n}d\alpha_{n}\, e^{- \frac{1}{2} \alpha_n^2} \delta
  (\alpha_1) \left| \sqrt{\frac{3}{5}} \alpha_0 - \frac{8}{\sqrt{35}} \alpha_2
  \right| \frac{1}{\Ndeltaphi} \,.
\end{equation}
The integrands of both Eqs.~\eqref{E:tostudy1} and~\eqref{E:tostudy2} have a dependence on $R$ which enters through $\Ndeltaphi$.

In the interacting theory, $S_b \sim \frac{1}{-\lambda} \gg 1$ while $\lambda_n
\sim 1$ so that the part of the path integral domain with $\alpha_n \lesssim 1$ dominates. Then we can approximate the Jacobian as
$\sqrt{\frac{3}{5}} \left( 2 \sqrt{S_b} + \alpha_0 \right) -
\frac{8}{\sqrt{35}} \alpha_2 \approx \sqrt{\frac{12S_b}{5}}$. If we
can justify ignoring the $\Ndeltaphi$ factor, the path integral reduces to the usual
expression used to compute the decay rate in $\lambda \phi^4$ QFT.

In the free theory, we also have $\alpha_n \lesssim 1$. There $S_b$ does not
appear and we can simply perform the integrals over $\alpha_0$ and $\alpha_2$ and
compare to what we would have gotten without collective coordinates, where the
$R$ integral is replaced by $\alpha_1$. The ratio of `collective coordinate integral' and the `non-collective coordinate integral' if we set
$\Ndeltaphi= 1$ is 
\begin{equation}
  \frac{\int \frac{d R}{R} \int d \alpha_0 \int d \alpha_2 e^{- \frac{1}{2}
  \alpha_0^2 - \frac{1}{2} \alpha_2^2} \left| \sqrt{\frac{3}{5}} \alpha_0 -
  \frac{8}{\sqrt{35}} \alpha_2 \right|}{\int d \alpha_0  d \alpha_1 d \alpha_2
  e^{- \frac{1}{2} \alpha_0^2- \frac{1}{2} \alpha_1^2 - \frac{1}{2}
  \alpha_2^2}} = \int \frac{d R}{R} \sqrt{\frac{17}{7 \pi^2}} \neq 1 \,.
\end{equation}
Since this ratio is not 1, we cannot expect $\Ndeltaphi = 1$ to be a good
approximation in the free theory. So if the standard computation for the interacting theory is correct, why can we drop $\Ndeltaphi$ when expanding around the bounce but not around the false vacuum?

To proceed, we have to understand the intersection number $\Ndeltaphi$. By analogy to the non-relativistic quantum mechanics case, $\Ndeltaphi$ counts
the number of solutions to $\langle \phi_d(x,R) | \phi \rangle = 0$ for a given
$\phi$ over the range of $R$. As in the quantum mechanics case, we need to
work in a basis other than the collective coordinate one to evaluate
$N_{\delta} [\phi]$. In the basis with fixed $\Rb$ as in Eq.~\eqref{R0form}, we find
\begin{equation}
  \big\langle \phi_d(x,R) \big| \phi(x) \big\rangle = 2 \sqrt{S_b} f_0(R,\Rb) + \sum_{n}c_n f_n (R, \Rb)
\end{equation}
where
\begin{equation}
  f_n (R, \Rb) = - \int d^4 x\, \phi_d (x, R) \Box \phi_n (x, \Rb) \,.
\end{equation}
For example, with $ z= R/\Rb$ we have
\begin{align}
  f_0 (z) &= \frac{\sqrt{15} z \left(-1-9 z^2+9 z^4+z^6-12 \left(z^4+z^2\right) \ln z\right)}{\left(z^2-1\right)^4}\\
  f_1 (z) &= -\frac{5 z \left(-1-46 (z^2- z^6)+z^8-12 z^2\left(3+z^2\right) \left(1+3 z^2\right)  \ln z\right)}{\left(z^2-1\right)^5} \,.
\end{align}

\begin{figure}
    \centering
    \includegraphics[scale=0.7]{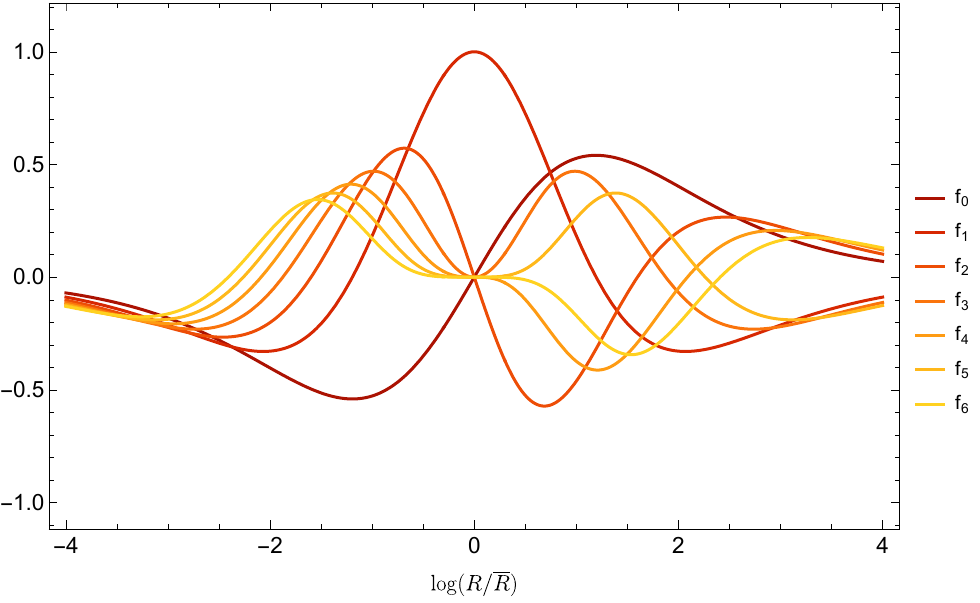}
    \caption{We depict the functions $f_n(R/\Rb) = \langle \phi_d(x,R)| \phi_n(x,\Rb)\rangle$. The collective coordinate $R$ is determined by the condition $2\sqrt{S_b}f_0(R/\Rb) + c_n f_n(R/\Rb) =0$ which may have multiple solutions.}
    \label{fig:fns}
\end{figure}

For the interacting theory, we can examine whether $\Ndeltaphi =1$ is a good approximation using the same approach as for the double-well potential in Section~\ref{sec:splitting}. Close to $R = \Rb$ we have
\begin{multline}
  \big\langle \phi_d(x,R)\big| \phi(x) \big\rangle = c_1 + \left( \sqrt{\frac{12 S_b}{5}} +
  \sqrt{\frac{3}{5}} c_0 - \frac{8}{\sqrt{35}} c_2 \right) \left(\frac{R-\Rb}{\Rb}\right) \\*
  + \frac{- 7 \sqrt{15} c_0 - 85 c_1 + 8 \sqrt{35} c_2 + 40 \sqrt{5} c_3-14\sqrt{15 S_b}}{70}
  \left(\frac{R-\Rb}{\Rb}\right)^2 + \cdots \,.
  \label{phidphi}
\end{multline}
Since $\sqrt{S_b} \gg 1$ we can solve
$\langle \phi_d | \phi \rangle=0$ as a series in $\frac{1}{\sqrt{S_b}}$. The leading terms are
\begin{equation}
  R = \Rb  \left( 1 - \sqrt{\frac{5}{12 S_b}} c_1 + \frac{14
  \sqrt{15} c_0 c_1 + 35 c_1^2 - 16 \sqrt{35} c_1 c_2}{168 S_b} + \cdots
  \right) \,. \label{smallR}
\end{equation}
Since $\lambda_n \approx 1$, the path integral
becomes suppressed if any $c_n \gtrsim 1$ on account of the $e^{- c_n^2 \lambda_n}$ terms. The only exception is the zero mode: excursions in the $c_1$ direction by arbitrary amounts only affect the integrand by an order 1 amount. From Eq.~\eqref{smallR} we see that there is always at least one solution with nonzero $c_1$.
To have multiple solutions the nonlinear terms need to become important. If we just consider the $c_n$ contributions, we require $c_n \gtrsim \sqrt{S_b} \gg 1$ and so the path integral is affected by an exponentially small amount (of order $e^{- c_1^2 \lambda_R} \approx e^{- S_b}$). The situation is akin to our quantum mechanical examples: the quantum fluctuations are too weak to significantly alter the classical trajectory. It remains to check the possibility of additional solutions due to large values of $c_1$. 

For large $c_1$, meaning $c_1 \gtrsim \sqrt{S_b}$, the expansion in Eq.~\eqref{smallR} is no longer useful. If we consider only paths of the form $\phi(x) = \phi_b(x,\Rb) + c_1 \phi_1(x,\Rb)$, then a solution to $\langle  \phi_d(x,R) | \phi(x)\rangle = 0$ requires
\begin{equation}
\frac{c_1}{\sqrt{S_b}} =-2 \,\frac{f_0(R,\Rb)}{f_1(R,\Rb)}
=
\vcenter{\hbox{
\begin{tikzpicture} 
\node[anchor=south west,inner sep=0] (image) at (0,0) {\includegraphics[scale=0.5]{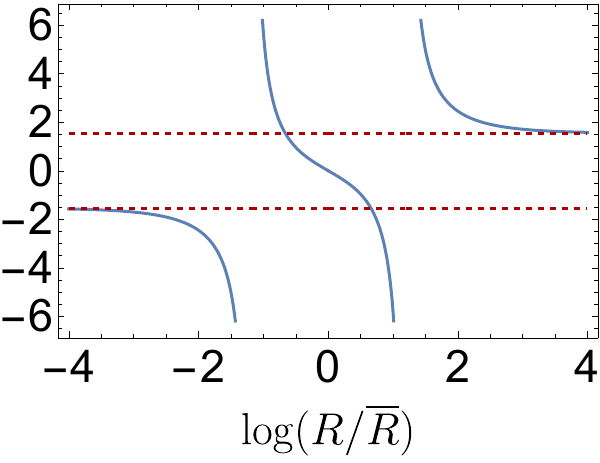}};
\node[scale=0.8,darkred] at (5.8,2.8) {$\sqrt{12/5}$};
\node[scale=0.8,darkred] at (5.8,2) {$-\sqrt{12/5}$};
\end{tikzpicture}
}}
\label{c1RR}
\end{equation}
For a given $c_1$ there is always at least one solution to the above equation. More precisely, since $f_1(R,\Rb)=0$ at $R/\Rb=x_c$ and $R/\Rb=1/x_c$ where $x_c \approx 0.303$, there will be one solution with $x_c \Rb < R < \frac{1}{x_c} \Rb$. However, if $|c_1/{\sqrt{S_b}}| > \sqrt{12/5}$, there will be a second solution. So here $1 \le \Ndeltaphi \le 2$, but we would like to determine whether the second solution is exponentially suppressed or not.

Although the action is exactly independent of the collective coordinate $R$, it is only independent of the coordinate $c_1$ to quadratic order. Expanding to higher orders there is a quartic term
\begin{equation}
    S[\phi_b + c_1 \phi_1] = S_b - \frac{5}{36 S_b} c_1^4  \,. \label{c14}
\end{equation}
The sign is negative since $\lambda < 0$; when expanding around the bounce to quadratic order, there is only one direction in which the potential decreases, but to quartic order it decreases in every direction (except the flat ones)! So we do not want to include
terms to quartic order in $c_1$. We do, however, want to include terms to quartic order in the collective coordinate $R$. If we did not include these terms, the action would not be independent of $R$. Large $R$ leads to a large factor of the volume of the dilatation group, but not an exponentially large factor from fluctuations in a direction of instability.

We are primarily interested in whether a field configuration has large quadratic action in our collective coordinates and not in the coordinates at fixed $\Rb$.  The $\alpha_n$ coordinates are given by
\begin{align}
    \alpha_n &= \big\langle \phi_b(x,\Rb) + c_1 \phi_1(x,\Rb) - \phi_b(x,R) \big| \phi_n(x,R)\big\rangle \label{alphanproj}\\
    &=g_n(R,\Rb) +c_1 f_n(R,\Rb)-2 \sqrt{S_b} \,\delta_{n0} 
\end{align}
where $g_n(R,\Rb) = \langle \phi_b(x,\Rb) | \phi_n(x,R)\rangle$ and we note that $g_1(R,\Rb) = f_0(R,\Rb)$. When $c_1 \ll \sqrt{S_b}$, we have $c_1 \approx \sqrt{\frac{12S_b}{5}} (R-\Rb)/\Rb$. In this regime $R$ is very close to $\Rb$ and all the $\alpha_n$ are small, namely $\alpha_n \ll \sqrt{S_b}$.
When $c_1 \gtrsim \sqrt{S_b}$ then all $\alpha_n \gtrsim \sqrt{S_b}$ as well. The corresponding field configurations, which include all of the $\Ndeltaphi=2$ field configurations with only $c_1$ turned on, are all suppressed. Other field configurations with $\Ndeltaphi>1$ would have various $c_n \gtrsim \sqrt{S_b}$ leading to $\alpha_n \gtrsim\sqrt{S_b}$ as well. 
Thus we can approximate $\Ndeltaphi$ as equal to $1$ up to exponentially small corrections in the path integral.

The free field theory behaves very differently, akin to the free quantum mechanics example.
In the free field theory, the $\sqrt{\frac{12 S_b}{5}} ( \frac{R -
\Rb}{\Rb} )$ term  in Eq.~\eqref{phidphi} is not present. Then the leading
solution in the regime $\frac{R - \Rb}{\Rb} \ll 1$ is
\begin{equation}
  R = \Rb  \left( 1 - \frac{7 \sqrt{5} c_1}{7 \sqrt{3} c_0 - 8
  \sqrt{7} c_2} + \cdots \right) \,.
\end{equation}
There is no small parameter in the free theory and no zero mode. So even with $c_n \lesssim 1$
there are multiple solutions to $\langle \phi_d | \phi \rangle
= 0$. That is, field configurations with $\Ndeltaphi \gtrsim 1$ are not suppressed and in fact $\Ndeltaphi$ can be arbitrarily large. Including these large values of $\Ndeltaphi$ must compensate for the unregulated $\int \frac{dR}{R}$ integral, akin to the free quantum mechanics case. Showing the cancellation explicitly seems extraordinarily challenging: collective coordinates around the bounce are very cumbersome to use for the free theory. 

An alternative approach to understand the exponential suppression when $\Ndeltaphi>1$ in the interacting theory, but not in the free theory, is to change coordinates to map directly to the double well. This approach is considered in Appendix~\ref{app:mapping}.

\subsection{Collective coordinates for translations}

In addition to the dilatation zero mode proportional to $\partial_R \phi_b$, the $\lambda \phi^4$ theory also has zero modes associated with translations which are linear combinations of the $\partial_\mu \phi_b$. When introducing collective coordinates for translations, one must also justify why the approximation $N_\delta=1$ is valid in the interacting theory but not for the free theory. The argument is essentially the same as for the double-well potential and for the dilatation mode: in the interacting theory paths with $N_\delta > 1$ give an exponentially suppressed contribution to the action. Because $\partial_\mu \phi_b$ breaks Lorentz invariance, or rather $SO(4)$ invariance in Euclidean signature, the modes which contribute to the projections $\langle \partial_\mu \phi_b(x) | \phi(x^\mu + x_0^\mu)\rangle$ are not so easy to identify.

To prove that the analysis for the translation modes proceeds as with the dilatation mode, the easiest approach is to make a coordinate transformation to map any particular direction in spacetime to $r$. This can be done because the theory is scale invariant and the classical theory has an exact $SO(5)$ symmetry as part of a larger conformal symmetry. To make the map explicit, we employ the conformal mapping with coordinates
\begin{equation}
    \eta^\mu = \frac{2x^\mu R}{R^2 + r^2},\quad \eta^5 = \frac{R^2-r^2}{R^2+r^2} \,.
\end{equation}
These new coordinates satisfy $\sum_{\mu = 1}^4 \eta_\mu^2 + \eta_5^2 = 1$, i.e.~they parameterize the 4-sphere. The inverse of this conformal mapping is the stereographic projection to Cartesian coordinates. 

On the 4-sphere, the integration volume is
\begin{equation}
  d^4 x = \left( \frac{2 R}{R^2 + r^2} \right)^{- 4} d \Omega_4 \,.
\end{equation}
If we also rescale the fields as
\begin{equation}
  \Phi (\eta) = \frac{r^2 + R^2}{2R} \phi (x) = \frac{R}{1 + \eta_5} \phi (x)\,, \label{Phimap}
\end{equation}
the action becomes
\begin{equation}
  \int d^4 x \left[ - \frac{1}{2} \phi \Box \phi + \frac{\lambda}{4} \phi^4
  \right] = \int d \Omega_4 \left[ \frac{1}{2} \Phi (- \vec{L}^2 + 2) \Phi +
  \frac{\lambda}{4} \Phi^4 \right],
\end{equation}
where $\vec{L}^2$ is the angular momentum operator. The equations of motion are now easily solved by a mode with $\vec{L}^2 \Phi = 0$, namely $\Phi_b = \sqrt{\frac{2}{-\lambda}}$.  Using Eq.~\eqref{Phimap} we find $\Phi_b =  \frac{r^2 +R^2}{2R} \phi_b$, as expected. 

The normal modes from expanding around $\Phi_b$ are spherical harmonics in 5 dimensions, which are identical to the conformally mapped fluctuations in Eq.~\eqref{normalmodes}. Indeed, it is through this conformal mapping that the exact eigenfunctions were first derived~\cite{Andreassen:2017rzq,Drummond:1979,McKane:1978}.
In these coordinates there is an exact $SO(5)$ symmetry among the 5 zero modes. We can therefore rotate any of the translation modes to the dilatation mode. Mapping back to flat space, the analysis of the path integral and $N_\delta$ proceeds identically to the discussion in Section~\ref{sec:dilatations}.

\section{Conclusions \label{sec:conc}}

In this paper we have provided a resolution to a long-standing puzzle involving collective coordinates. Collective coordinates are a way to factor out an integral over the volume of a symmetry group from the path integral and still exploit a saddle point approximation in the non-symmetry directions. The puzzle is how collective coordinates can be consistently introduced through a change of coordinates, independent of the action. Doing so appears to produce correct results in interacting theories where the symmetry volume is expected, but not in free theories, where it is unexpected. The resolution to this puzzle is that the change of coordinates is allowed and valid, but it is multi-valued: a given path $x(t)$ in quantum mechanics may be described by multiple values of the collective coordinates $t_0$. We quantified this multi-valuedness with the intersection number $\Ndelta$: it counts the number of times two paths $X(t)$ and $x(t)$ intersect as one is shifted by the collective coordinate over the symmetry volume. The correct way to introduce a collective coordinate is therefore with a resolution of the identity of the form
\begin{equation}
    1 = \int d t_0 \frac{1}{\Ndelta}\,
  \delta\Big[\langle X (t+t_0) |x (t)\rangle \Big] \,
  \Big| \langle X (t+t_0) | \dot{x}  (t)
  \rangle \Big|  \,.
\end{equation}
Without the factor of $\Ndelta$, the path integral can give the wrong answer.

To call out and rectify the mistreatment of collective coordinates we have explored examples in quantum mechanics and quantum field theory. For the splitting between the two lowest energy levels in the symmetric double well potential in quantum mechanics, the $\Ndelta$ factor counts the number of times a path crosses the potential barrier. Because each crossing gives additional suppression by $e^{-S_I}$ where $S_I$ is the instanton action, we argued that one can approximate $\Ndelta = 1$ up to exponentially small corrections.  We made special effort to regulate the theory in finite time and carefully take the limit of a combination of the twisted and untwisted partition function to extract the energy splitting. While the final result is well-known through the WKB approximation, most path integral treatments involve unjustified approximations. We have attempted to provide a complete path integral
treatment of the splitting where each step is under explicit control.

In the interacting theories we have explored here, namely the double well in quantum mechanics and interacting scalar quantum field theory, we found that the intersection number could be approximated as 1. In the free theories, the intersection number cannot be approximated as 1 and in fact it can be arbitrarily large, as needed to compensate for the volume factor which arises in collective coordinates.

It would be interesting to explore cases intermediate between these two extremes. An example is the instanton-anti-instanton pair in QCD as a function of separation. This configuration interpolates between the vacuum at zero separation and a saddle point of the action at infinite separation. The separation distance is a quasi-collective coordinate~\cite{Balitsky:1985in, Balitsky:1986qn,Yung:1987zp}, becoming a flat direction at infinite separation and a unexceptional normal mode fluctuation at zero separation. In the dilute instanton gas approximation, the dependence of the action on the separation distance is neglected, so that the separation is treated as a collective coordinate. Thus the techniques developed here could provide insights into the validity of the dilute instanton gas approximation.
Although QCD is a significant step up in complexity from the examples considered here, it is of direct phenomenological relevance and is certainly worth examining in more detail.

\subsection*{Acknowledgements}
We would like to thank G.~Dunne, M.~Mari\~no, M.~Serone, and J.~Stout for valuable discussions.  M.D.S.~and A.B.~are supported in part by the U.S.~Department of Energy under contract DE-SC0013607.  J.C.~is supported by a Junior Fellowship from the Harvard Society of Fellows.

\appendix
\section{Metric approach to Jacobians in the path integral \label{app:metric}}
In this Appendix we explain an approach for computing Jacobians arising from changes of coordinates in path integrals.  To this end, we review how to use Riemannian geometry to manipulate the path integral measure. For concreteness we consider a quantum mechanical path integral for a particle in $d$ spatial dimensions, although our arguments here readily generalize to field theory. 

The path integral measure and its normalization depend on a metric on the space of paths.  If $K$ is a positive-definite operator, we can begin by defining
an inner product
\begin{equation}
\label{E:innerproductK}
    \big\langle x(t) \big| y(t) \big\rangle_K = \int dt \,x(t) \cdot K \cdot y(t)\,.
\end{equation}
It is most common to then choose a basis of functions $x_n(t)$ that are orthonormal with respect to this inner product. If we write $x(t) = \sum_n c_n x_n(t)$  where $\langle x_m | x_n \rangle = \delta_{mn}$ then the path integral measure is
\begin{equation}
    {\mathcal D}x = \prod_n d c_n\,.
\end{equation}
Factors of $2\pi$ or an overall normalization ${\mathcal N}$ can be optionally absorbed into the measure; these will drop out of physical quantities calculated by ratios of path integrals.

Now we consider a set of coordinates which are not orthonormal. If we have any set of coordinates $\{a_n\}$ which can parametrize a general path $x(t)$ then we can define a distance element between paths as
\begin{equation}
    ds^2 =  \sum_{m,n} G_{m n}(a) da_m d a_n=
   \sum_{m,n} \Bigg\langle \frac{\partial x}{\partial a_m} \Bigg|  \frac{\partial x}{\partial a_n} \Bigg\rangle \,da_m \, da_n\,.
\end{equation}
In this case the path integral measure becomes
\begin{equation}
\label{E:measure0}
    {\mathcal D}x = \sqrt{\det G }\prod_n d a_n\,,
\end{equation}
which is the analog of the measure $\sqrt{\det g}\,d^d x$ in Riemannian geometry.

When we compute correlation functions, we divide by a partition function normalization computed using the same measure in Eq.~\eqref{E:measure0}.  We observe that correlation functions are independent of the choice of positive-definite $K$ in Eq.~\eqref{E:innerproductK}, since changing $K$ to $K'$ affects the square root determinant multiplicatively by $\sqrt{\frac{\det K'}{\det K}}$ which cancels out the same factor in the partition function normalization.

Let us show how these more abstract ideas manifest in two examples.
\\ \\
\textbf{Example 1:} First, suppose we consider paths $x : [0,T] \to \mathbb{R}$ with $x(0) = x_i$ and $x(T) = x_f$.  The space of such paths can be parameterized by
\begin{align}
x(t) = x_i + \frac{t}{T}(x_f - x_i) + \sum_{n = 1}^\infty c_n \, \sqrt{\frac{2}{T}}\sin\!\left(\frac{\pi n}{T}\,t\right)
\end{align}
with coordinates $\{c_n\}$.  Taking $K=1$,  corresponding to the ordinary `flat' $L^2$ inner product, the metric is
\begin{equation}
    G_{m n}=
    \Big\langle \frac{\partial x}{\partial c_m} \Big|  \frac{\partial x}{\partial c_n} \Big\rangle 
    = \frac{2}{T}\int_0^T dt\, \sin\!\left(\frac{\pi m}{T}\,t\right) \,\sin\!\left(\frac{\pi n}{T}\,t\right) = \delta_{mn} 
\end{equation}
and so 
\begin{align}
ds^2 = \sum_{n} dc_n^2\,.
\end{align}
For this metric $\sqrt{\det G} = 1$
and the path integral measure is simply 
\begin{align}
\mathcal{D}x = \prod_{n}\,dc_n\,.
\end{align}
\\
\textbf{Example 2:} Now consider coordinates for paths around some instanton $x_I(t)$, namely $x(t) = x_I(t) + \sum_{n = 0}^\infty a_n\,x_n(t)$.  If $\langle x_m(t)| x_n(t) \rangle = \delta_{mn}$, then, by the same arguments as in the first example above, the path integral measure can be expressed in coordinates $\{a_n\}$ as $\prod_{n = 0}^\infty da_n$.  If $x_0(t)$ is a time translation zero mode of the action of our theory, then it is instead convenient to work with an expansion $x(t) = x_I(t + t_0) + \sum_{n = 1}^\infty \alpha_n\,x_n(t + t_0)$ in coordinates $\{t_0, \alpha_{n > 0}\}$.  In these coordinates, the metric is
\begin{equation}
ds^2 = \langle \dot{x}(t)|\dot{x}(t) \rangle\,dt_0^2 + 2 \sum_{n = 1}^\infty\langle \dot{x}(t) |x_n(t + t_0)\rangle\,dt_0 \,d\alpha_n + \sum_{n = 1}^\infty d\alpha_n^2
\end{equation}
and so 
\begin{equation}
\sqrt{\det G} =\left(\langle \dot{x}(t)|\dot{x}(t) \rangle -  \sum_{n = 1}^\infty\langle \dot{x}(t) |x_n(t + t_0)\rangle \langle x_n(t + t_0)| \dot{x}(t)\rangle \right)^{1/2}\,.
\end{equation}
Since $\sum_{n = 0}^\infty\langle a(t) |x_n(t + t_0)\rangle \langle x_n(t + t_0)| b(t)\rangle = \langle a(t)| b(t)\rangle$ by completeness of $\{x_{n \geq 0}(t + t_0)\}$ for fixed $t_0$, the above simplifies to
\begin{equation}
\sqrt{\det G} = \left|\langle x_0(t + t_0)|\dot{x}(t)\rangle\right| = \left|\langle x_0(t)| \dot{x}_I(t) + \sum_{n = 1}^\infty \alpha_n \dot{x}_n(t)\rangle\right|\,.
\end{equation}
Thus in the $\{t_0, \alpha_{n > 0}\}$ coordinates the path integral measure is 
\begin{equation}
\mathcal{D}x = dt_0 \prod_{m = 1}^\infty d\alpha_m\, \left|\langle x_0(t)| \dot{x}_I(t)\rangle + \sum_{n = 1}^\infty \alpha_n \langle x_0(t) | \dot{x}_n(t)\rangle\right|
\end{equation}
which agrees with Eq.~\eqref{jformtext}.
Of course, we also have to be careful about the change in the domain of path integration coming from our change in coordinates.  The modification of the domain and subtleties with multi-valued coordinate changes are more readily understood using the methods in the main text.

\section{Mapping QFT to the double well \label{app:mapping}}

In the quantum mechanics case, we saw that $N_\delta$ counted the number of times a particular path crossed $x=0$. This led to a simple physical argument for why paths with $N_\delta > 1$ were exponentially suppressed in the interacting theory where the potential is high at $x=0$, but unsuppressed in the free theory where $x=0$ is not special in any sense. In the scalar quantum field theory we justified why $N_\delta=1$ is appropriate for the interacting theory, but did not provide a physical interpretation of $N_\delta$. We can do so by mapping the theory to quantum mechanics. This mapping provides an alternative way of thinking about the bounces in QFT.

To map QFT to QM we follow~\cite{Yung:1987zp, Khoze:219734} and use a conformal mapping to turn the dilatation direction into
time. Let us take $r = R e^{\omega \tau}$ with $\omega$ an arbitrary frequency, and consider spherically-symmetric
field configurations of the form
\begin{equation}
\label{E:heq1}
  \phi (r) = e^{- \omega \tau} h (\tau) \,.
\end{equation}
Then the action of $\lambda \phi^4$ theory evaluated on such field configurations is
\begin{equation}
  S [\phi] = \frac{2\pi^2R^2}{\omega} \int d \tau \left[ \frac{1}{2}
  (\partial_{\tau} h)^2 + \frac{1}{2} \omega^2 h^2 + R^2 \omega^2
  \frac{\lambda}{4} h^4 \right] \,.
\end{equation}
The equations of motion are then $h'' (\tau) - V' (h) = 0$ where $V (h) =
\frac{1}{2} \omega^2 h^2 + R^2 \omega^2 \frac{\lambda}{4} h^4$. Recalling that $\lambda <
0$, the above is an inverted double-well potential. The bounce
is written as
\begin{equation}
  h_b (\tau) = \sqrt{\frac{2}{- \lambda}} \frac{1}{R \cosh (\omega \tau)} \,.
\end{equation}
The above satisfies the equation of motion of a particle in an inverted inverted
double well, starting at the metastable local maximum at
$h = 0$. The particle rolls slowly off of this hump starting at $\tau = - \infty$, 
bounces off a wall, and ends up back on top of the hump at $\tau = \infty$.
The flat direction is associated with the time at the center of the bounce.
The normal modes which had satisfied $\Box \phi_n = \lambda_n V'' (\phi_b)
\phi_n$ now satisfy 
\begin{equation}
  [\partial_{\tau}^2 - \omega^2] h_n = - \lambda_n \frac{6 \omega^2}{\cosh^2 (\omega \tau)}
  h_n \,.
\end{equation}
However, there is a better basis we can use. Expanding around $h_b (\tau)$ the
quadratic fluctuations satisfy
\begin{equation}
  \left[ \partial_{\tau}^2 - \omega^2 + \frac{6\omega^2}{\cosh^2 (\omega \tau)} \right]
  \eta_n = \hat{\lambda}_n \eta_n \,,
\end{equation}
which reproduces the P\"{o}schl-Teller potential with $\lambda = 2$. There are thus two bound states
and a continuum. The $h_n$ and $\eta_h$ bases are different, except for the
dilatation zero mode
\begin{equation}
  h_d (\tau) = \eta_d (\tau) = \frac{\sqrt{5}}{4 \pi R} \frac{\sinh (\omega
  \tau)}{\cosh^2 (\omega \tau)}
\end{equation}
which has the same form in both bases (it has $\hat{\lambda}_1 = 0$ and
$\lambda_1 = 1$ so that $\frac{\lambda_1 - 1}{\lambda_1} = 0$ as in Eq.
\eqref{eq:squadbounce}).

The zero condition $\langle \phi_d (x,R) | \phi (x) \rangle = 0$ becomes
$\langle \eta_1 (\tau-\tau_0) |h (\tau) \rangle = 0$ after the mapping to the double well, and can be
studied using $\eta$ modes quite similarly to the way we studied the double
well in Section~\ref{sec:doublewell}. We write
\begin{equation}
  h (\tau) = b_0 \eta_0 (\tau) + b_1 \eta_1 (\tau)+ \sum_{n\geq 2} b_n
  (\tau) \eta_n (\tau)
\end{equation}
where $\eta_n$ are the continuum modes and $\eta_0, \eta_1$ are
the bound states. As with the double well, the bound states have support for
$\tau \sim \omega^{- 1}$ while the continuum modes must be regulated with
$-T < \tau < T$, or equivalently $R e^{- \omega T} < r < R e^{\omega
T}$. The projection that needs to be studied now involves the collective coordinate $\tau_0$ and is
\begin{equation}
  \langle \eta_1 (\tau - \tau_0) |h (\tau) \rangle = 0 \,.
\end{equation}
In the regime $\tau_0 \ll \omega^{- 1}$ we reduce to the case where we can
trade $\tau_0$ for $b_1$. In the regime $\omega^{- 1} \ll \tau_0 \ll T$ the
two bound states have support only near $\tau = 0$ and in particular
\begin{equation}
  \eta_1 (\tau - \tau_0) \approx \delta' (\tau - \tau_0), \quad \omega^{- 1}
  \ll \tau_0 \ll T\,,
\end{equation}
so that the zero overlap condition becomes $h' (\tau_0) = 0$. Thus $\tau_0$ is the
time when $h (\tau)$ stops and $N_{\delta} [h]$ counts the number of times $h$
stops.

Plugging~\eqref{E:heq1} into the action of the free theory, the potential is
$V_0 (h) = \frac{1}{2} \omega^2 h^2$ which is just a stable quadratic
potential. Thus the $h$ actions for the free and interacting theories are similar, and the action of
a given path in the interacting theory is always strictly lower than the
action in the free theory. Nevertheless, since the saddle point we expand around has large action in the interacting theory, perturbations around this saddle must be large to slow the particle down so that it stops multiple times. Such large quantum fluctuations are exponentially suppressed.  In the free theory, the saddle point is just $h=0$ and the particle can stop multiple times without much penalty. Hence in the full quantum field theory setting, $N_d[\phi]=1$ is a good approximation for the interacting theory but not for the free theory.

\bibliographystyle{JHEP}
\bibliography{refs}
\end{document}